\documentclass[review]{elsarticle}

\usepackage{epstopdf}
\usepackage{subfigure}
\usepackage{color,amsmath,amsfonts}
\usepackage{algorithmic,algorithm}

\newcommand{\sigmaa}{\sigma_\mathrm{a}}
\newcommand{\sigmas}{\sigma_\mathrm{s}}
\newcommand{\tol}{\epsilon}

\newcommand{\change}[1]{\textcolor{black}{#1}}

\begin{document}


\bibliographystyle{elsarticle-num}
\title{Data-Driven Acceleration of Thermal Radiation Transfer Calculations with the Dynamic Mode Decomposition and a Sequential Singular Value Decomposition}
\author[mymainaddress]{Ryan G. McClarren\corref{mycorrespondingauthor}}
\ead{rmcclarr@nd.edu}
\cortext[mycorrespondingauthor]{Corresponding author}

\address[mymainaddress]{University of Notre Dame, Dept.~of Aerospace \& Mechanical Engineering, 365 Fitzpatrick Hall, Notre Dame, Indiana, USA}
\author[tmainaddress]{Terry S. Haut}
\ead{haut3@llnl.gov}

\address[tmainaddress]{Lawrence Livermore National Laboratory, Livermore, California, USA}

\begin{abstract}
We present a method for accelerating discrete ordinates radiative transfer calculations for radiative transfer. Our method works with nonlinear positivity fixes, in contrast to most acceleration schemes.  The method is based on the dynamic mode decomposition (DMD) and using a sequence of rank-one updates to compute the singular value decomposition needed for DMD.  Using a sequential method allows us to automatically determine the number of solution vectors to include in the DMD acceleration.  We present results for slab geometry discrete ordinates calculations with the standard temperature linearization. Compared with positive source iteration, our results demonstrate that our acceleration method reduces the number of transport sweeps required to solve the problem by a factor of about 3 on a standard diffusive Marshak wave problem, \change{a factor of several thousand on a cooling problem where the effective scattering ratio approaches unity},  and a factor of 20 improvement in a realistic,  multimaterial radiating shock problem.
\end{abstract}

\begin{keyword}
Radiative transfer; discrete ordinates method; dynamic mode decomposition; acceleration methods;
\end{keyword}

\maketitle

\section{Introduction}
The dynamic mode decomposition (DMD) \cite{Schmid:2010ee,Schmid:2010hh} is a data-driven method for understanding the spectral properties of an operator. It relies solely on a sequence of vectors generated by an operator and requires no knowledge of the operator. \change{In the computational fluid dynamics community it has been used} for understanding the properties of flows \cite{Tu:2013dj} and for comparing simulation and experiment \cite{Schmid:2010ee}.  For neutron transport problems it was introduced as a technique to estimate time-eigenvalues \cite{mcclarren2018calculating}, for creating reduced order models \cite{hardy2019dynamic}, to understand stability \cite{di2020dynamic}, and for accelerating power iterations for $k$-eigenvalue problems \cite{roberts2019acceleration}. In this work we turn to the problem of accelerating discrete ordinates solutions to radiative transfer problems (primarily x-ray radiative transfer for time-dependent high-energy density physics applications).

In radiative transfer problems positive solutions, by which we mean positive radiation densities, are essential due to the coupling of the radiation transport equation to an equation for the material temperature.  Negative densities can lead to negative material temperatures that are both nonphysical and cause instabilities. Moreover, many numerical methods based on high-order representations of the solution \cite{Mathews:1999uv} or different angular discretizations \cite{McClarren:2008p976} can lead to negative radiation densities.

Methods to remove the negative solutions that can arise from these problems have been presented over the years.  The zero-and-rescale fix \cite{Hamilton:2009uo,yee2019quadratic} sets any negative values to zero and scales other unknowns to conserve particles.   This method has been shown to be effective, but it does not preserve certain moments of the transport equation.  The consistent set-to-zero method (CSZ) \cite{Maginot:2010vx} addresses this problem by solving a local nonlinear equation to remove nonlinearities.  Other attempts to address negative solutions are the exponential discontinuous method \cite{Walters:1996iu} and the positive spherical harmonics method \cite{HauckMac2009, Laiu:2016jx}.

All of these methods to remove negative solutions (with the exception of the exponential discontinuous scheme) render the solution of radiation transport equation nonlinear. Positive source iteration (a form of nonlinear Richardson iteration) can still be used, but can be arbitrarily slow to converge on diffusive problems \cite{Adams:2001vt}. Nevertheless, the nonlinear nature of the solution technique implies that standard acceleration techniques based on linear problems such as diffusion synthetic acceleration (DSA) \cite{Adams:2001vt} and preconditioned GMRES \cite{Warsa:2004p1085} can no longer be used.  There have been attempts to derive acceleration methods based on Jacobian-free Newton-Krylov \cite{Fichtl:2017fd} and nonlinear acceleration through a quasi-diffusion approach \cite{yee2019quadratic}. 

In this paper we propose to use a simple acceleration based on the dynamic mode decomposition.  We use DMD to estimate the slowest decaying error modes in positive source iteration, and then estimate the solution. Because DMD is a data-driven method, it is simple to implement.  DMD relies on the computation of a singular value decomposition (SVD) of a data matrix containing the solution for the scalar intensity (i.e., the scalar flux) at several iterations.  To alleviate the expense of this decomposition we employ a sequential algorithm to perform the SVD that estimates the SVD using rank-one updates. Additionally, because we use a sequential algorithm, we can automatically determine the number of iterations to include in the DMD update. The inclusion of a sequential SVD and the automatic selection of the number of iterations are the two key improvements over our preliminary work presented at a conference \cite{mcclarren2018acceleration} \change{and  a similar approach for linear problems given by Andersson and Eriksson \cite{andersson2014novel}.}

While data-driven methods may seem outside the normal toolkit of  particle transport research, it is worth noting that Krylov methods such as GMRES can be thought of as data-driven because they do not require the knowledge of the matrix, rather just the action of the matrix.  We also believe that the explosion of data being generated in the computational sciences will be another rowel to investigate more of these kinds of methods.

\change{
In this work we compare a DMD acceleration technique based on a sequential SVD to positive source iteration and demonstrate significant reduction in the number of iterations required to converge. Though other acceleration techniques have been proposed \cite{Fichtl:2017fd,yee2019quadratic}, and many others are possible---including nonlinear GMRES \cite{washio1997krylov} and nonlinear Krylov acceleration \cite{till2013application}---a thorough, fair comparison of these methods is outside the scope of this work. Such a fair comparison would need to employ the methods in the same code base and utilize the latest implementations of all the constituent parts such as solvers, preconditioners, parallel strategies, etc. and would be an excellent contribution to state of knowledge for future work.
}

This paper is organized as follows.  In section 2 we introduce the dynamic mode decomposition and its properties.  We then in section 3 discuss the gray, discrete ordinates radiative transfer equations and the discontinuous Galerkin discretization of those equations using Bernstein polynomials. Section 4 gives the standard, unaccelerated positive source iteration method, before we present the DMD acceleration for that method in Section 5. Section 6 gives numerical results followed by conclusions and future work in section 7.

\section{The dynamic mode decomposition}

\label{sec:DMD}
Here we discuss the properties of the dynamic mode decomposition (DMD) for approximating an operator based on information from the action of the operator.  A thorough treatment of the theory of this decomposition can be found in \cite{Schmid:2010ee,Schmid:2010hh,Schmid:2011ec,Tu:2013dj}.

We consider a sequence vectors $y_k$ that are related by the application of an operator $A$:
\begin{equation}\label{eq:vec_relation}
    y_{k+1} = A y_{k}.
\end{equation}
The vectors $y_k \in \mathbb{R}^N$, $A$ is an operator of size $N\times N$, and $k=0,\dots,K$. The vectors $y_k$ could come from a discretized PDE, experimental measurements, sensor readings, etc.  As we will see, knowledge of $A$ is not required; only the $y_k$ need to be known.

To find the DMD decomposition, we append the vectors into a {\em data matrices} of size $N\times K$ as
  \begin{equation} Y_+ = \begin{pmatrix}   | & | & & | \\ y_1 & y_2 & \dots &y_K \\  | & | & & |  \end{pmatrix}
  \qquad Y_- = \begin{pmatrix}   | & | & & | \\ y_0 & y_1 & \dots &y_{K-1} \\  | & | & & |  \end{pmatrix}.\end{equation}
  With the data matrices \change{one} can write Eq.~\eqref{eq:vec_relation} as \begin{equation}\label{eq:mat_relation} Y_+ = A Y_-.\end{equation}

We then take the thin singular value decomposition (SVD) of $Y_-$ to write 
\begin{equation} Y_- = U \Sigma V^\mathrm{T},\end{equation}
  where $U$ is a $N \times K$ orthogonal matrix, $\Sigma$ is a diagonal $K\times K$ matrix with non-negative entries on the diagonal, and $V$ is a $K \times K$ orthogonal matrix.  The matrix $U$ has columns that form an orthonormal basis for the row space of $Y_-  \subset \mathbb{R}^N$. In the case when there are only $r < K$ nonzero singular values, we use the compact SVD where $U$ is $N \times r$, $\Sigma$ is $r \times r$, and $V$ is $r \times K$.

We substitute the SVD of $Y_-$ into Eq.~\eqref{eq:mat_relation}, to get 
\begin{equation} Y_+ = A U \Sigma V^\mathrm{T},\end{equation}
and then we use the orthonormality properties of $V$ and $U$, and the fact that $\Sigma$ is a diagonal matrix with non-zero entries to write
\begin{equation}\label{eq:Atilde}
   \tilde{A} \equiv U^\mathrm{T} A U = U^\mathrm{T} Y_+ V \Sigma^{-1}.
   \end{equation}
The matrix $\tilde{A}$ is an $r \times r$ matrix that is a rank $r$ approximation to $A$, where $r$ is the number of nonzero singular values in the SVD of $Y_-$. Notice in Eq.~\eqref{eq:Atilde} that $\tilde{A}$ can be formed using only the data matrices and no knowledge of $A$.

The dynamic modes of $A$ are determined from the eigenvalues of $\tilde{A}$. This requires solving an $r\times r$ eigenvalue problem. If $(\lambda, w)$ are eigenvalue/eigenvector pairs of $\tilde{A}$ then
\begin{equation}\label{eq:modeDef}
    \varphi = \frac{1}{\lambda}Y_+ V \Sigma^{-1} w
\end{equation}
are the $r$ dynamic modes of $A$.  The mode with the largest value of $\lambda$ is said to be the dominant mode.

One of the properties of DMD is that the dynamic modes found will depend on the modes excited by the data. For instance, if $y_0$ is an eigenvector of $A$, then only one mode will be excited.  This property was used previously in time-eigenvalue problems in neutron transport to find eigenmodes important to the evolution of an experiment \cite{mcclarren2018calculating}.

Before moving on, we point out that despite the fact that DMD is derived as a linear method, it has been shown that DMD can be applied to nonlinear operators.  In particular DMD will find an approximation to the Koopman operator for the nonlinear update \cite{Tu:2013dj}. This will allow us to use DMD on nonlinear solution techniques for the radiative transfer equations.
\section{The Gray, Discrete Ordinates Radiative Transfer Equations}
We will apply DMD to accelerate the solution to gray, radiative transfer calculations using discrete ordinates (S$_N$). The S$_N$ equations of thermal radiative transfer in slab geometry the high-energy density regime are given by
\begin{subequations}\label{eq:radtran}
\begin{equation}
    \frac{1}{c} \frac{\partial I_n}{\partial t} + \mu_n\frac{\partial I_n}{\partial x} + \sigmaa(x,t,T) I_n = 
    \frac{1}{2}\sigmaa a c T^4 + \frac{1}{2} Q(x,t),
\end{equation}
\begin{equation}
    \frac{\partial e}{\partial t}  =  \sigmaa(x,t,T)(\phi - a c T^4),
\end{equation}
\begin{equation}
    \phi(x,t) = \sum_{n=1}^N w_n I_n.
\end{equation}
\end{subequations}
Here $x$ [cm] is the spatial variable, $t$ [ns] is the time variable, $w_n$ and $\mu_n$ are the weights and abscissas of a quadrature rule over the range $(-1,1)$, $I_n(x,t)$ [GJ/(cm$^2\cdot$s$\cdot$steradian)] is the specific intensity of radiation in the quadrature direction $n$, $\phi(x,t)$  [GJ/(cm$^2\cdot$s)] is the scalar intensity, $T(x,t)$ [keV] is the material temperature, and $e(T)$ [GJ] is the internal energy density of the material related to $T$ via a known equation of state. Additionally, $c \approx 30$ [cm/ns] is the speed of light, $a = 0.01372$ [GJ/(cm$^3\cdot$keV$^{4}$)], $\sigmaa(x,t,T)$ [cm$^{-1}$] is the absorption opacity, and $Q(x,t)$ is a known, prescribed source. For quadrature rules we apply Gauss-Legendre quadrature rules of even order. 

The boundary conditions for Eq.~\eqref{eq:radtran} prescribe an incoming intensity on the boundary:
\begin{equation}
    I_n(0,t) = g_n(t)\quad \mu_n > 0, \qquad I_n(X,t) = h_n(t)\quad \mu_n < 0,
\end{equation}
where $g_n$ and $h_n$ are known functions of time and $X$ is the right boundary of the problem domain. Initial conditions specify $I_n(x,0)$ throughout the problem.

For time discretization we use the backward Euler method with a linearization of the nonlinear temperature term. We write the solution at time $t = m\Delta t$ using the superscript $m$: $I_n^m(x) = I(x,m\Delta t)$. The semi-discrete equations are \cite{mcc2008} 
\begin{subequations}\label{eq:radtran_semi}
\begin{equation}
    \mu_n\frac{\partial I_n^{m+1}}{\partial x} + \sigma^* I^{m+1}_n = 
    \frac{1}{2}\left(\sigmas^m \phi^{m+1} +  \sigmaa^m f a c (T^m)^4\right) + Q^*,
\end{equation}
\begin{equation}
    e^{m+1}  =  e^{m} + \Delta t \sigmaa^m (\phi^{m+1} - a c (T^m)^4),
    \end{equation}
\end{subequations}
where $\sigma^* = \sigmaa^m + (c\Delta t)^{-1}$, $Q^* = Q^{m+1}/2 + (c\Delta t)^{-1}$, $\sigmas = (1-f) \sigmaa^{m}$ is the effective scattering term, and the factor $f$ is defined as
\begin{equation}\label{eq:fleck}
    f(x,t,T) = \frac{1}{1+\beta c \sigmaa \Delta t}, \qquad \beta = \frac{4a}{C_\mathrm{v}}
\end{equation}
with $C_\mathrm{v}$ the heat capacity at constant volume for the material.  It is also useful to define a radiation temperature as $T_\mathrm{r} = \sqrt[4]{\phi/(ac)}$.

The system in \eqref{eq:radtran_semi} is a quasi-steady transport problem to which we apply  a discontinuous Galerkin finite element method in space using the Bernstein polynomials as a basis \cite{yee2019quadratic,haut2019efficient}.  The resulting equations are
\begin{subequations}\label{eq:radtran_discrete}
\begin{equation}\label{eq:dis_trans}
    (\mu_n \mathbf{G} + \mathbf{F}_n + \mathbf{M}^*)\mathbf{I}_n^{m+1} =\frac{1}{2}\left(\mathbf{M}_\mathrm{s}\boldsymbol{\phi}^{m+1} +\mathbf{M}_\mathrm{a} ac(\mathbf{T}^m)^4\right) + \mathbf{Q}^{*},
\end{equation}
\begin{equation}\label{eq:dis_mat}
    \mathbf{M}_e(\mathbf{e}^{m+1}-\mathbf{e}^{m})  =  \mathbf{M}_\mathrm{a} (\boldsymbol{\phi}^{m+1} - a c (\mathbf{T}^{m})^4),
\end{equation}
\end{subequations}
where the superscript $m$ denotes a time level, $\mu_n\mathbf{G} + \mathbf{F}_n$ is the upwinded representation of the derivative term, $\mathbf{M}^*, \mathbf{M}_\mathrm{s},$ and  $\mathbf{M}_\mathrm{a}$ are the mass matrices associated with the $\sigma^*, \sigmas,$ and $\sigmaa$ terms, respectively. The vectors $\mathbf{I}_n^m, \boldsymbol{\phi}^m, \mathbf{T}^m,$ and $\mathbf{Q}^*$ are vectors that contain the coefficients of the finite element representations of the intensity, scalar intensity, temperature, and source. 

The system in \eqref{eq:radtran_discrete} can be advanced in time by solving Eq.~\eqref{eq:dis_trans} and then evaluating the material internal energy update in Eq.~\eqref{eq:dis_mat}. However, the addition of the effective scattering term on the RHS of Eq.~\eqref{eq:dis_trans} couples all of the $N$ quadrature directions together. 

\section{Positive Source Iteration Method}
The matrices on the LHS of Eq.~\eqref{eq:dis_trans} can be written in block lower-triangular form \cite{Warsa:2004p1085}. Therefore, we can perform the following iterative procedure to find $\mathbf{I}_n^{m+1}$
\begin{equation}
    \label{eq:SI}
    \left.\mathbf{I}_n^{m+1}\right|_{k+1} =(\mu_n \mathbf{G} + \mathbf{F}_n + \mathbf{M}^*)^{-1}\left[\frac{1}{2}\left(\left.\mathbf{M}_\mathrm{s}\boldsymbol{\phi}^{m+1}\right|_k +\mathbf{M}_\mathrm{a} ac(\mathbf{T}^m)^4\right) + \mathbf{Q}^{*}\right].
\end{equation}
Here we denote the $k$th iteration of a quantity as $\left.(\cdot)\right|_k$. The application of the inverse of the the lower triangular operator $(\mu_n \mathbf{G} + \mathbf{F}_n + \mathbf{M}^*)$ is known as a transport sweep: it involves moving information for a particular direction $n$ across the problem domain. Note that if we take the quadrature sum of both sides of Eq.~\eqref{eq:SI} we get an update in terms of the scalar intensity only:
\begin{equation}
    \label{eq:SI_scalar}
    \left.\boldsymbol{\phi}^{m+1}\right|_{k+1} =\mathbf{D}(\mu_n \mathbf{G} + \mathbf{F}_n + \mathbf{M}^*)^{-1}\left[\frac{1}{2}\left(\left.\mathbf{M}_\mathrm{s}\boldsymbol{\phi}^{m+1}\right|_k +\mathbf{M}_\mathrm{a} ac(\mathbf{T}^m)^4\right) + \mathbf{Q}^{*}\right],
\end{equation}
where $\mathbf{D}$ represents the quadrature sum $\sum_{n=1}^N w_n$ operator.

The iteration scheme in Eq.~\eqref{eq:SI_scalar} can be very slow to converge when $f \rightarrow 0$ and/or $\Delta t \rightarrow \infty$. In this scenario the discrete equations have no absorption of radiation, leading to the iterations having a spectral radius approaching unity \cite{Adams:2001vt}. It has been shown that the iterations can be accelerated by using a diffusion correction, called diffusion synthetic acceleration, and by ``wrapping'' the iterations in a Krylov solver and preconditioning the solver \cite{Warsa:2004p1085}.

\subsection{Positivity Fixes}

Physically, the specific intensity is a phase-space density, and as such it should be non-negative.  Nevertheless, it is known that solutions to discrete ordinates problems can give negative solutions \cite{yee2019quadratic,Mathews}. This is particularly vexing in radiative transfer problems because negative intensities can lead to negative temperatures \cite{McClarren:2008p976} that can cause issues with evaluating material properties.

To address this issue we use the zero-and-rescale fix \cite{yee2019quadratic} to impose positivity on the intensities  in our calculations. This is a nonlinear method that during the transport sweep monitors the solution during the sweep. If one of the coefficients is negative, this implies that the finite element representation will have negative values. Therefore, we zero out any negative coefficients and rescale the other coefficients local to a zone to conserve the total intensity of the solution locally. Using transport sweeps with the zero-and-rescale fix is a form of nonlinear Richardson iteration.

The addition of this nonlinear fix renders acceleration techniques such as diffusion synthetic acceleration and preconditioned GMRES impotent as these techniques require a linear iterative strategy. Recently, Yee, et al.~ showed that this nonlinear fix could be accommodated in a nonlinear quasi-diffusion iteration \cite{yee2019quadratic}.  Here we will show how DMD can be used to handle this type of nonlinearity as well.

\section{DMD Acceleration}
In this section we show how DMD can be applied to source iteration using a sequential SVD. To begin we write Eq.~\eqref{eq:SI_scalar} in the following shorthand:
\begin{equation}
    \label{eq:shorthandSI} 
    y_{k+1} = A y_{k} + {b}
\end{equation}
where $y_{k+1} = \left.\boldsymbol{\phi}^{m+1}\right|_{k+1}$, and 
\begin{equation}
A = \frac{1}{2}\mathbf{D}(\mu_n \mathbf{G} + \mathbf{F}_n + \mathbf{M}^*)^{-1}\mathbf{M}_\mathrm{s},
\end{equation}
\begin{equation}
b = \mathbf{D}(\mu_n \mathbf{G} + \mathbf{F}_n + \mathbf{M}^*)^{-1}\left[\mathbf{M}_\mathrm{a} \frac{ac}{2}(\mathbf{T}^m)^4 + \mathbf{Q}^{*}\right].
\end{equation}

By substituting in the converged solutions, we can see that Eq.~\eqref{eq:shorthandSI} is an iterative procedure for solving
\begin{equation}\label{eq:shorthandSI_y}
    (I-A)y = b,
\end{equation}
where $I$ is the identity operator. Also, if we subtract successive iterations we get the following relationship for the difference between iterations:
\begin{equation}
y_{k + 1} - y_k = A (y_k - y_{k-1}).
\end{equation}
It is this relationship that we will use \change{with} DMD to formulate an approximation to $A$. We define data matrices to contain the differences between iterations
\begin{equation}\label{eq:YplusDiff}
Y_+ = [y_2-y_1, y_3-y_2, \dots, y_{K+1} - y_{K}],
\end{equation}
\begin{equation}\label{eq:YminusDiff}
Y_- = [y_1-y_0, y_2-y_1, \dots, y_{K} - y_{K-1}].
\end{equation}
These are each $N \times K$ matrices, where $N$ is the number of spatial degrees of freedom.
As before we define an approximate $A$ as the $K\times K$ matrix:
\begin{equation}\tilde{A} = U^\mathrm{T} A U = U^\mathrm{T} Y_+ V \Sigma^{-1}.\end{equation}
 We can use $\tilde{A}$ to construct the operator $(I-\tilde{A})^{-1}$ and use this to approximate the solution. 
 
Using Eq.~\eqref{eq:shorthandSI_y} we can write the difference between the solution and the $K$th iteration as
\begin{align}
    \nonumber (I-A) (y - y_{K}) &= b - (I-A)y_K\\
    \nonumber &= b - y_{K} + (y_{K+1}-b)\\
    \label{eq:Kiterate} &= y_{K+1} - y_{K}.
\end{align}
Next, we define $\Delta z$ as the length $K$ vector that satisfies
\begin{equation}
    y-y_{K} = U \Delta z,
\end{equation}
and substitute this into the LHS of Eq.~\eqref{eq:Kiterate} and left multiply by $U^\mathrm{T}$ to get
\begin{equation}
    \label{eq:DMD_Update} (I - \tilde{A}) \Delta z = U^\mathrm{T} (y_{K+1}-y_{K}).
\end{equation}
This is a linear system of size $r \leq K$ where $r$ is the number of non-singular values in the SVD of $Y_-$. We solve this system and approximate the solution as $y \approx y_{K} + U \Delta z$.

This algorithm uses the changes between iterations to estimate the operator $A$ that governs the iterative change.  We then, in effect, use this approximated operator to extrapolate the solution to convergence.  This update requires taking $K+1$ iterations of source iteration, the computation of an SVD, and the solution of a small linear system.

\subsection{Sequential SVD and Automatic DMD}
In the previous algorithm, we needed to compute $K+1$ iterations in addition to the SVD.  However, choosing how many iterations is not obvious. Additionally, the SVD will require $O(NK^2)$ operations to compute where $N$ is the number of spatial degrees of freedom. To address both of these problems we use a sequential SVD generated by rank-one updates.

\change{Brand \cite{brand2006fast} presented an algorithm for taking the SVD of a data matrix where the elements in the matrix were generated sequentially. The resulting cost of the SVD is then $O(NKr)$ where $r$ is the rank of the SVD. This algorithm was then used by Choi et al.~\cite{choi2019space} to develop reduced order models for particle transport problems.} We use this approach to build up the SVD using successive source iterations and determine, based on the results, when $K$ is large enough. The function defined in \cite{choi2019space} {\tt  incrementalSVD} takes as inputs a new column vector, $u$, a tolerance for linear dependence $\epsilon_\mathrm{SVD}$, a minimum size for a singular value $\epsilon_{SV}$, the current singular value decomposition $U,\, S,\, V$ and the column index of the vector, $k$. Thus, we write a call of the incremental SVD as  {\tt  incrementalSVD}{$(u, \epsilon_\mathrm{SVD}, \epsilon_{SV},U, S, V, k)$}.

\subsection{Acceleration Algorithm}
We specify the algorithm for applying automatic DMD with sequential SVD in Algorithm \ref{al:AutomaticDMD}.  This algorithm uses the incremental SVD function defined in Algorithm 2 of \cite{choi2019space}.  Our automatic DMD acceleration takes successive source iterations to build up the data matrices defined in Eqs.~\eqref{eq:YplusDiff} and \eqref{eq:YminusDiff}. After two source iterations, we have enough data to start applying the acceleration.  We compute a new value of the estimated solution based on the approximation $\tilde{A}$ as in Eq.~\eqref{eq:DMD_Update}. We continue making approximations until either the maximum number of iterations is reached or until we find that the two source iterations did not add to the rank of $\tilde{A}$. This stopping criteria is used because the data indicates that further source iterations are not improving the approximation $\tilde{A}$. 

There are other particularities of the automatic DMD acceleration that we point out here.  Firstly, we remove small singular values from the SVD of $Y_-$. This is done to remove singular values that are unimportant and could add numerical noise to the update. Additionally, we do not compute an update based on DMD if there are eigenvalues of $\tilde{A}$ with a magnitude larger than one (c.f.~line 18 of the algorithm).  This is because these large eigenvalues could allow the solution to diverge in the update.
 \begin{algorithm}
  \caption{Automatic DMD Acceleration}\label{al:AutomaticDMD}
   $[\boldsymbol{\phi}] =$ AutomaticDMDAcceleration($A$, $b$, $\left.\boldsymbol{\phi}\right|_0$, $\tol$)\\
   \textbf{Input:} Sweep operators $A$ and $b$, initial guess $\left.\boldsymbol{\phi}\right|_0$, maximum iterations $K$, current residual estimate $\tol$\\
   \textbf{Output:} Approximate solution $\boldsymbol{\phi}$
    \begin{algorithmic}[1]
    \STATE  tmpOld $= \left.\boldsymbol{\phi}\right|_0$
    \STATE $Y_+ =[]$, $Y_- = []$
    \STATE $U =[]$, $\Sigma = []$, $V = []$
    \FOR{k = 1 \TO K+1}
    \STATE tmp$ = A\cdot$tmpOld$+b$, i.e., Perform a sweep
    \STATE $\boldsymbol{\phi} = $ tmp
    \IF{$k < K+1$}
    \STATE Append $\Delta_k=(\mathrm{tmp}-\mathrm{tmpOld})$ to $Y_-$
    \ENDIF
    \IF{$k > 1$}
    \STATE Append $\Delta_k=(\mathrm{tmp}-\mathrm{tmpOld})$ to $Y_+$
    \STATE $[U,\Sigma,V,r] =$  \textbf{incrementalSVD}($\Delta_k$, $\tol \cdot 10^{-14}$, $\tol \cdot 10^{-14}$, $U$,
    $S$, $V$, $k-1$)
    \ENDIF
    \IF{$k > 2$}
    \STATE Remove singular values from $\Sigma$ less than $\tol \times 10^{-6}$ of the trace of $\Sigma$
    \STATE $\tilde{A} = U^\mathrm{T} Y_+ V \Sigma^{-1} $
    \STATE Compute eigenvalues of $\tilde{A}$ as $\lambda_k$
    \IF{$\max_k(|\lambda_k|) < 1$}
    \STATE Solve $(I - \tilde{A}) \Delta z = U^\mathrm{T} \Delta_k $ for $\Delta z$
    \STATE $\boldsymbol{\phi} = \mathrm{tmpOld} + U\Delta z$
    \ELSE
    \STATE Not enough iterations to estimate $\tilde{A}$, continue
    \ENDIF
    \ENDIF
    
    \STATE tmpOld = tmp
    \IF{$k > r + 2$}
    \STATE Exit For Loop
    \ENDIF
    \ENDFOR
    \RETURN $\boldsymbol{\phi}^{m+1}$
    \end{algorithmic}
 \end{algorithm}

 The time update using automatic DMD acceleration is shown in Algorithm \ref{al:RadDMD}. When we apply the automatic DMD acceleration to compute a time update, we add $J$ additional source iterations outside the DMD acceleration.  This is done to damp any high-frequency errors introduced by the DMD acceleration.  In practice we typically use $J=2$ or $3$. In Algorithm \ref{al:RadDMD} we check for convergence in the source iterations outside the DMD acceleration step.  In practice we also check for convergence in the DMD acceleration function to save on iterations, but this detail is omitted from our listing for clarity in the algorithms.
  \begin{algorithm}
  \caption{Radiative Transfer Time Step Update with DMD}\label{al:RadDMD}
  [ $\mathbf{I}_n$, $\boldsymbol{\phi}$, $\mathbf{e} $,  $\mathbf{T}$] = RadStep($\boldsymbol{\phi}^{m},$ $\mathbf{I}_n^m$, $\mathbf{T}^m$, $\mathbf{e}^m$, $J$, $K$, $\tol_2$, $\tol_\infty$,\dots)\\
   \textbf{Input:} Previous solutions $\boldsymbol{\phi}^{m},$ $\mathbf{I}_n^m$, $\mathbf{T}^m$, and $\mathbf{e}^m$, material properties, quadrature rule, number of extra iterations $J$ and maximum DMD iterations $K$, $\tol_2$ and $\tol_\infty$ as the $L_2$ and $L_\infty$ tolerances\\
   \textbf{Output:} Solutions at time level $m+1$: $\boldsymbol{\phi},$ $\mathbf{I}_n$, $\mathbf{T}$, and $\mathbf{e}$.
    \begin{algorithmic}[1]
    \STATE  Compute $b=\mathbf{D}(\mu_n \mathbf{G} + \mathbf{F}_n + \mathbf{M}^*)^{-1}\left[\mathbf{M}_\mathrm{a} \frac{ac}{2}(\mathbf{T}^m)^4 + \mathbf{Q}^{*}\right]$
    \STATE $\left.\boldsymbol{\phi}\right|_0 = \boldsymbol{\phi}^{m}$
    \WHILE{Not Converged}
    \STATE \{Apply Source Iteration $J$ times\}
    \FOR{j=1 \TO J}
        \STATE $\left.\boldsymbol{\phi}\right|_j = A\left.\boldsymbol{\phi}\right|_{j-1} + b$
        \STATE change $=\|\left.\boldsymbol{\phi}\right|_j - \left.\boldsymbol{\phi}\right|_{j-1}\|_2$
        \IF{change $ < \tol_2$ \AND $\|\left.\boldsymbol{\phi}\right|_j - \left.\boldsymbol{\phi}\right|_{j-1}\|_\infty < \tol_\infty$}
        \STATE \{The iterations are converged\}
            \STATE $\mathbf{I}_n   =(\mu_n \mathbf{G} + \mathbf{F}_n + \mathbf{M}^*)^{-1}\left[\frac{1}{2}\left(\left.\mathbf{M}_\mathrm{s}\boldsymbol{\phi}\right|_j +\mathbf{M}_\mathrm{a} ac(\mathbf{T}^m)^4\right) + \mathbf{Q}^{*}\right]$
            \STATE $\mathbf{e} = \mathbf{e}^{m} + \mathbf{M}_\mathrm{e}^{-1}\mathbf{M}_\mathrm{a} (\boldsymbol{\phi} - a c (\mathbf{T}^{m})^4)$
            \STATE Compute $\mathbf{T}$ by inverting the equation of state at $\mathbf{e}$
            \RETURN $\mathbf{I}_n$, $\left.\boldsymbol{\phi}\right|_j$, $\mathbf{e} $, and $\mathbf{T}$.
        \ENDIF
    \ENDFOR
    \STATE \{Apply DMD Acceleration\}
    \STATE $\left.\boldsymbol{\phi}\right|_0 =$ AutomaticDMDAcceleration($A$, $b$,$\left.\boldsymbol{\phi}\right|_J$, change)
    \ENDWHILE
    \end{algorithmic}
 \end{algorithm}
 
 To compute a time step for the radiative transfer solver, the storage requirements for the radiation variables are
 \begin{itemize}
     \item Two angular flux vectors for the previous and current angular flux,
     \item The data matrices, $Y_+$ and $Y_-$, each of size $N \times k$ where $N$ is the number of spatial degrees of freedom, and $k \leq K$ is the number of iterations required in the DMD acceleration step.
 \end{itemize}
 The number of iterations (transport sweeps) required for convergence will be the sum of the iterations outside the DMD acceleration step and those required in the DMD update. For comparison with standard source iteration, we use the number of transport sweeps as our metric for efficiency.
 
 For the nonlinear zero-and-rescale fix, we apply that nonlinearity during the transport sweep, i.e., in the application of $\mathbf{D}(\mu_n \mathbf{G} + \mathbf{F}_n + \mathbf{M}^*)^{-1}$. This is not explicitly called out in Algorithm \ref{al:RadDMD}, but will be understood in our results.
 
\section{Numerical Results}
\subsection{Diffusive Marshak wave}
 To demonstrate the effectiveness of our acceleration strategy, we consider a standard, diffusive Marshak wave problem \cite{Petschek, Lane:2013ku,mcc2008}. We use $\sigma = 300 T^{-3}$, and an equation of state given by $e = C_\mathrm{v} T$ with $C_\mathrm{v} = 0.3$ GJ/(keV$\cdot$cm$^{-3}$); there is no source in the problem.  The initial conditions are $T(x,0) = 0.001$ keV and $\phi(x,0) = acT(x,0)^4$. The domain has an incoming boundary condition of $g_n = ac/2$ at $x=0$ and no incoming radiation at the right edge of the domain.
 
 \change{This problem will require positivity fixes, primarily near the wave front. In our numerical calculations we find that anywhere from 0.1 to 1.5\% of mesh zones visited during the calculation (i.e., up to 2\% of the product $I N M$ where $I$ is the number of zones, $N$ is the number of angles, and $M$ is the number of iterations). }
 
\begin{figure}
     \centering
     \includegraphics[width=0.8\textwidth]{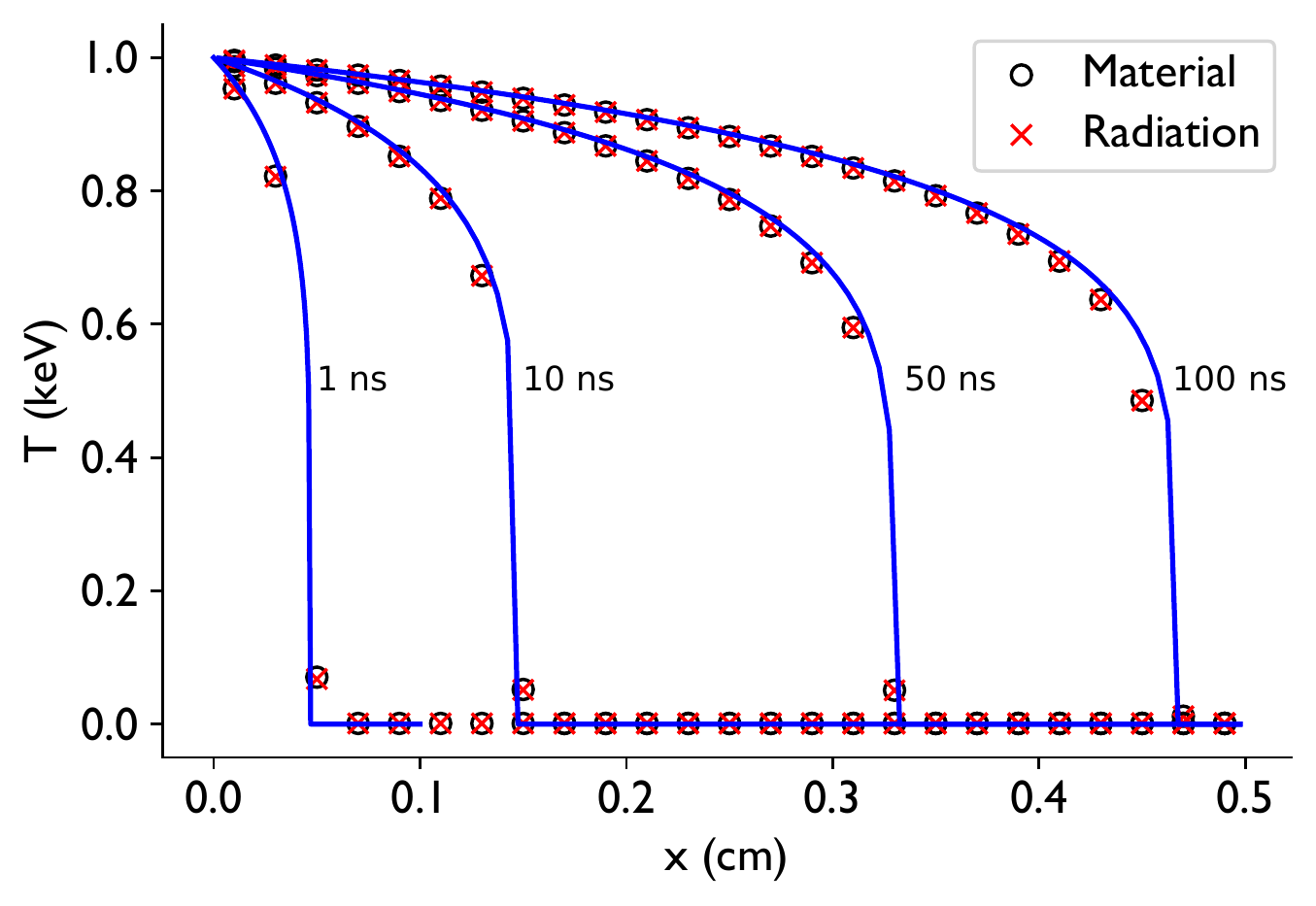}
     \caption{Comparison of S$_8$ solutions obtained with DMD and semi-analytic diffusion solution for the Marshak wave problem with at 4 different times. The S$_8$ solutions used $\Delta t = 0.01$ ns, zone sizes of $0.02$ cm, and a cubic polynomial basis.}
     \label{fig:marshak_compare}
 \end{figure}

We run the problem with different values of the FEM expansion order, number of spatial zones, and with a time step size of $\Delta t = 0.01$ ns and S$_8$ Gauss-Legendre quadrature. In Figure \ref{fig:marshak_compare} results from the DMD solution with cubic elements of size $0.02$ cm are compared with the semi-analytic diffusion solution.  The source iteration solutions are identical on the scale of the figure to the DMD solutions and are, therefore, not shown. 
 In the figure we see that the S$_8$ solution agrees with the semi-analytic solution except near the wavefront, as has been previously observed in comparisons with the diffusion solution \cite{Morel:2006cq,McClarren:2008it,SmedleyStevenson:2015gt}.

\begin{figure}
    \centering
    \includegraphics[width=0.8\textwidth]{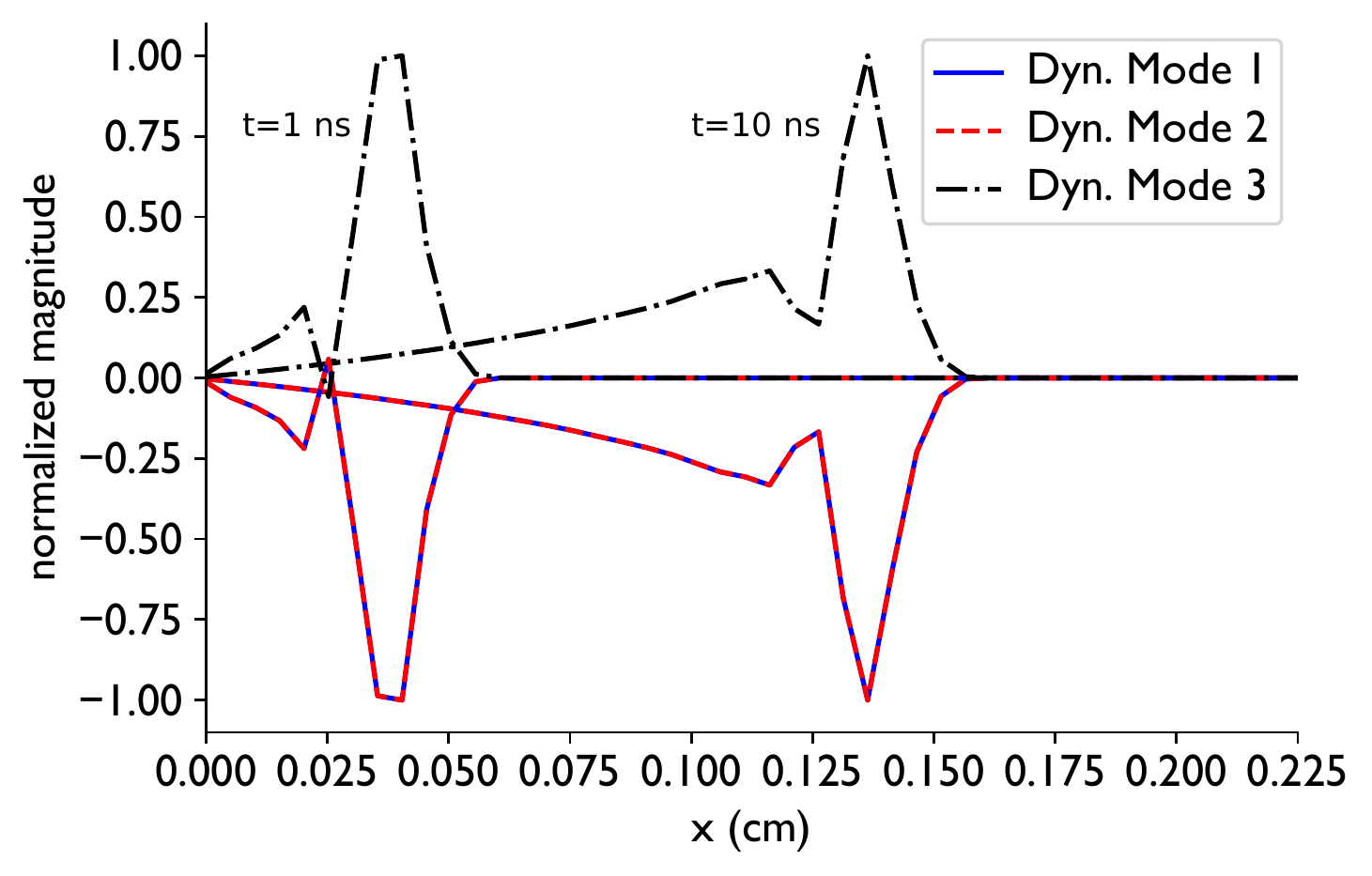}
    \caption{Three most dominant dynamic modes found during the time steps at times $t=1$ and $10$ ns. The eigenvalues of $\tilde{A}$ at the different times are $\{8.56\times 10^{-3}, 2.33\times 10^{-4}, 2.21\times10^{-6}\}$ and  $\{1.54\times 10^{-3}, 4.71\times 10^{-5}, 2.81\times10^{-7}\}$.}
    \label{fig:marshdynamic}
\end{figure}

For this problem we inspect the three dominant dynamic modes in the update of $\phi$ found from the first application of DMD acceleration in the step at $t=1$ and $10$ ns. These dominant modes will be estimates of the slowest decaying error modes from source iteration. In Figure \ref{fig:marshdynamic} we plot  modes as calculated by Eq.~\eqref{eq:modeDef}; we normalize the modes by dividing by the maximum magnitude in the mode. Note that these magnitude are only defined up to a factor of $\pm1$.  From Figure \ref{fig:marshdynamic} we see that the dominant modes highlight the wavefront and the heated region behind it. This is expected because most of the change in the solution is occurring at the wavefront. Also, it is here that positivity preservation is needed. The fact that the three modes have nearly the same shape indicates that there are several, similar error modes that are slowly decaying.

\subsubsection{Comparison with Source Iteration}

\begin{figure}
\subfigure[$\Delta t= 0.005$]{\includegraphics[width=0.49\textwidth]{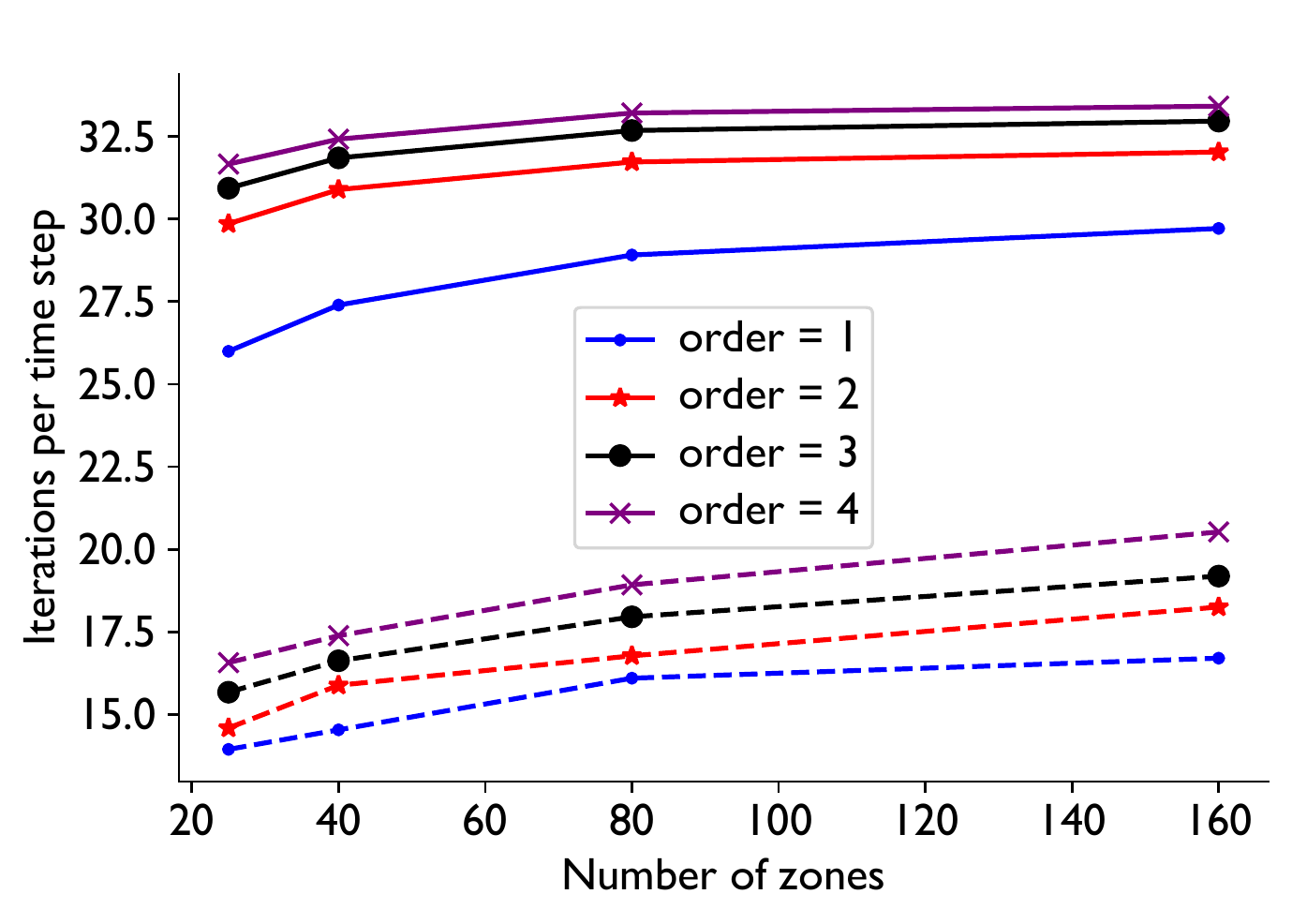}}
\subfigure[$\Delta t= 0.01$]{\includegraphics[width=0.49\textwidth]{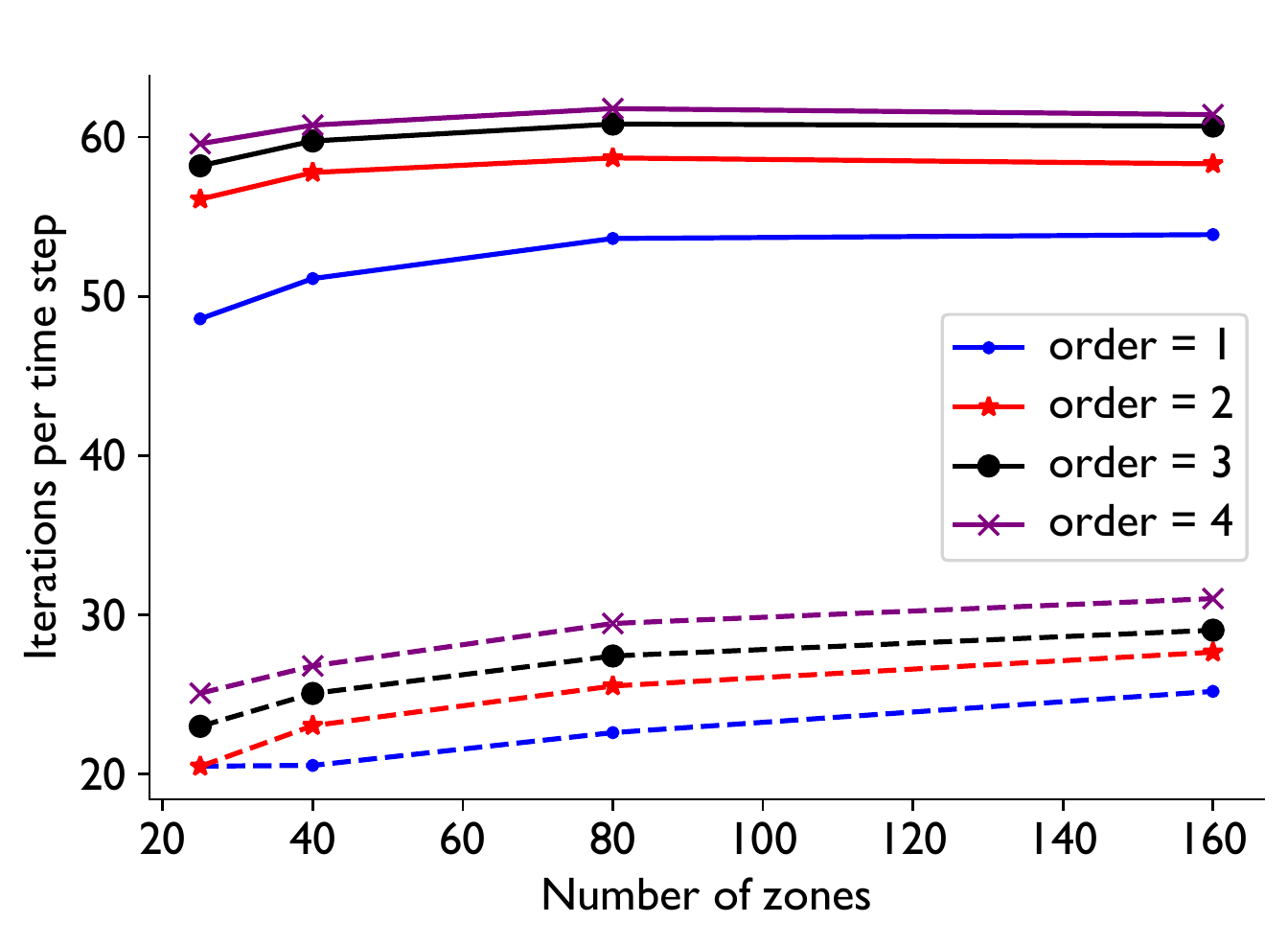}}
\subfigure[$\Delta t= 0.02$]{\includegraphics[width=0.49\textwidth]{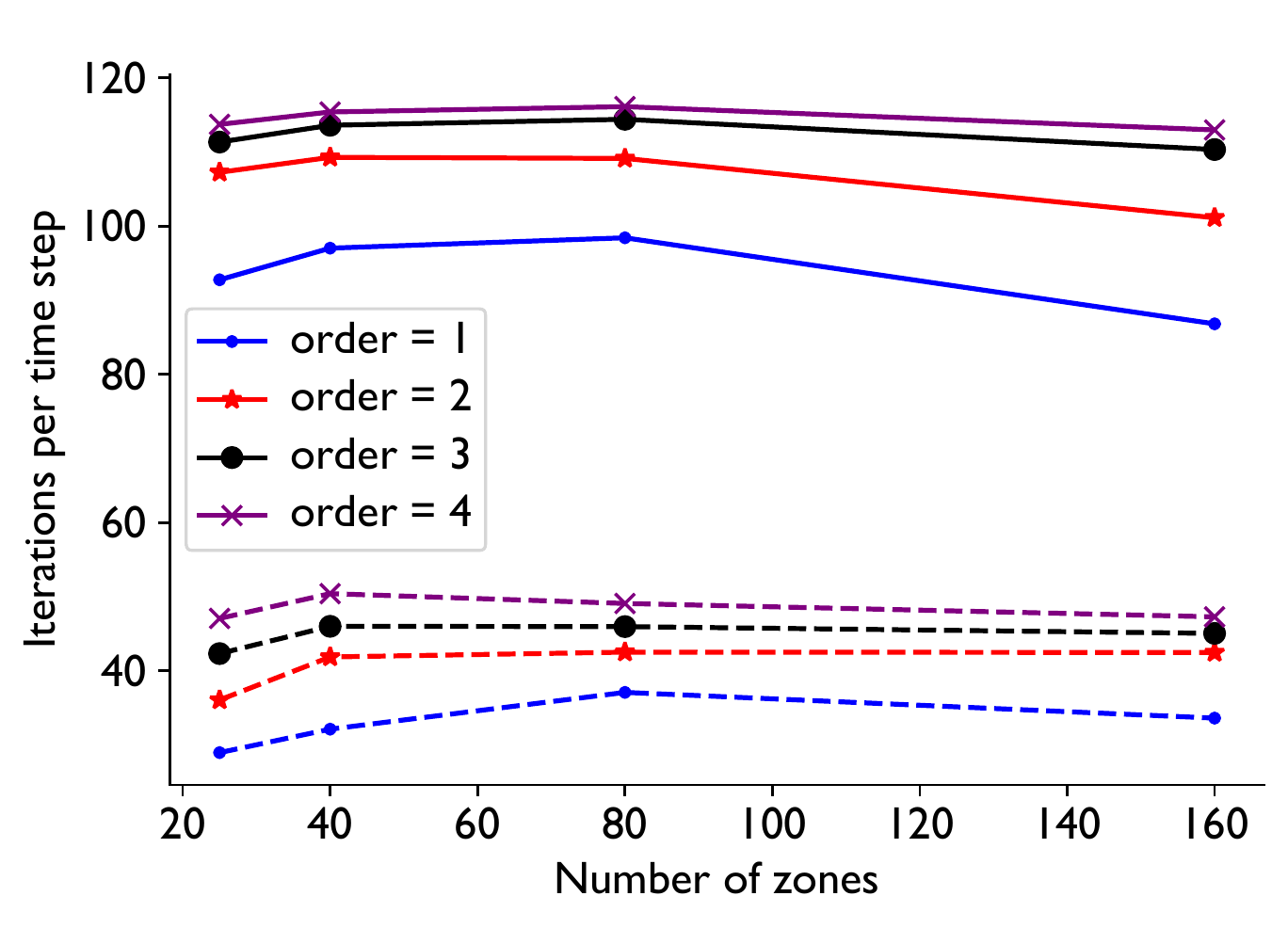}}
	\caption{Number of iterations per time step for the Marshak wave problem using  several different time step sizes, number of zones, and expansion order of the FEM solution. The problem was run until a final time of $t=10$ ns. Solid lines denote positive source iteration results; dashed lines are for DMD-accelerated calculations.}
	\label{fig:iterations_marshak}
\end{figure}

\change{To compare the efficacy of our positive, automatic DMD acceleration with standard positive source iteration we vary the time step size, number of zones, and order of the FEM expansion.  In Figure \ref{fig:iterations_marshak} the average number of iterations, that is the number of transport sweeps, per time step is shown. We note that for this problem the positivity fix we utilize is needed as the solution near the wavefront can become negative when the fix is not applied.}
 
For this Marshak wave problem, the time step size is a proxy for the scattering ratio, $\sigmas/\sigma^*$. Using the definition of these quantities and the material properties of this problem, we find the scattering ratio simplifies to 
\begin{align}
    \frac{\sigmas}{\sigma^*} &= \frac{(1-f) \sigmaa}{\sigmaa + \frac{1}{c\Delta t}} \\ \nonumber
    &\approx \frac{1600 \Delta t }{1600 \Delta t +1}.
\end{align}
For $\Delta t = 0.005, 0.01, $ and $0.02$ ns the corresponding scattering ratios are 0.8889, 0.9412, 0.9697, respectively. 

We solve the Marshak wave problem until a final time of $t=10$ ns using a variety of spatial resolutions, time steps, and finite element expansion orders. We report the number of iterations (i.e., transport sweeps) required to solve the problem to the final time in Figure \ref{fig:iterations_marshak}.  From the figure we see that the DMD-accelerated solutions require significantly fewer iterations than positive source iteration, fewer than half as many iterations. The difference between the required number of iterations gets larger as the scattering ratio increases to about a 40\% reduction when $\Delta t = 0.02$. For this problem we do not increase the time step further as this causes overheating due to the large time step and actually makes the medium behave {\em less} diffusive.

\change{Because DMD is a data-driven acceleration technique, there is no guarantee that changing the number of degrees of freedom will not change the behavior of the iterative convergence.} We notice that as the order of the finite element expansion increases the number of iterations required also increases. This was consistent in the source iteration and DMD-accelerated results. We also observe a slight increase in the number of iterations required in DMD as the number of zones increases in the $\Delta t = 0.005$ and 0.01 ns cases.  The number of iterations required in the $\Delta t = 0.02$ ns case, however, decreases as the number of zones increases above 40.
\change{These changes in the number of iterations required as a function of the number of spatial zones and the finite element order track those of previous results for linear problems using a piecewise constant discretization \cite{mcclarren2018acceleration}.}

\subsection{Cooling problem}
\change{
While the Marshak wave a is a standard problem in radiative transfer, we are limited in how difficult we can make it from a convergence point of view: the ratio $\sigmas/\sigma*$ can only be made so large. This is due to the fact that  large time steps can make the problem nonphysically overheat, which makes the wave travel faster, and make the problem easier.}

\change{
To address this we contrived a test problem of a slab initially at a uniform temperature of 0.5 keV with a radiation temperature of 0.45 keV surrounded by vacuum.  The slab is 1 cm thick and has a $\sigmaa = \sigma_0 T^{-m}$ where $m = 0$ or $3$. The problem is run for a single time step of 0.01 ns; the spatial discretization has 50 zones and order 3 polynomials. We will adjust $\sigma_0$ from 10 to 10$^6$ to make the problem more difficult.  Though this problem will not require positivity fixes, it is indicative of a worst case scenario.
}
\begin{figure}
    \centering
    \includegraphics[width=0.8\textwidth]{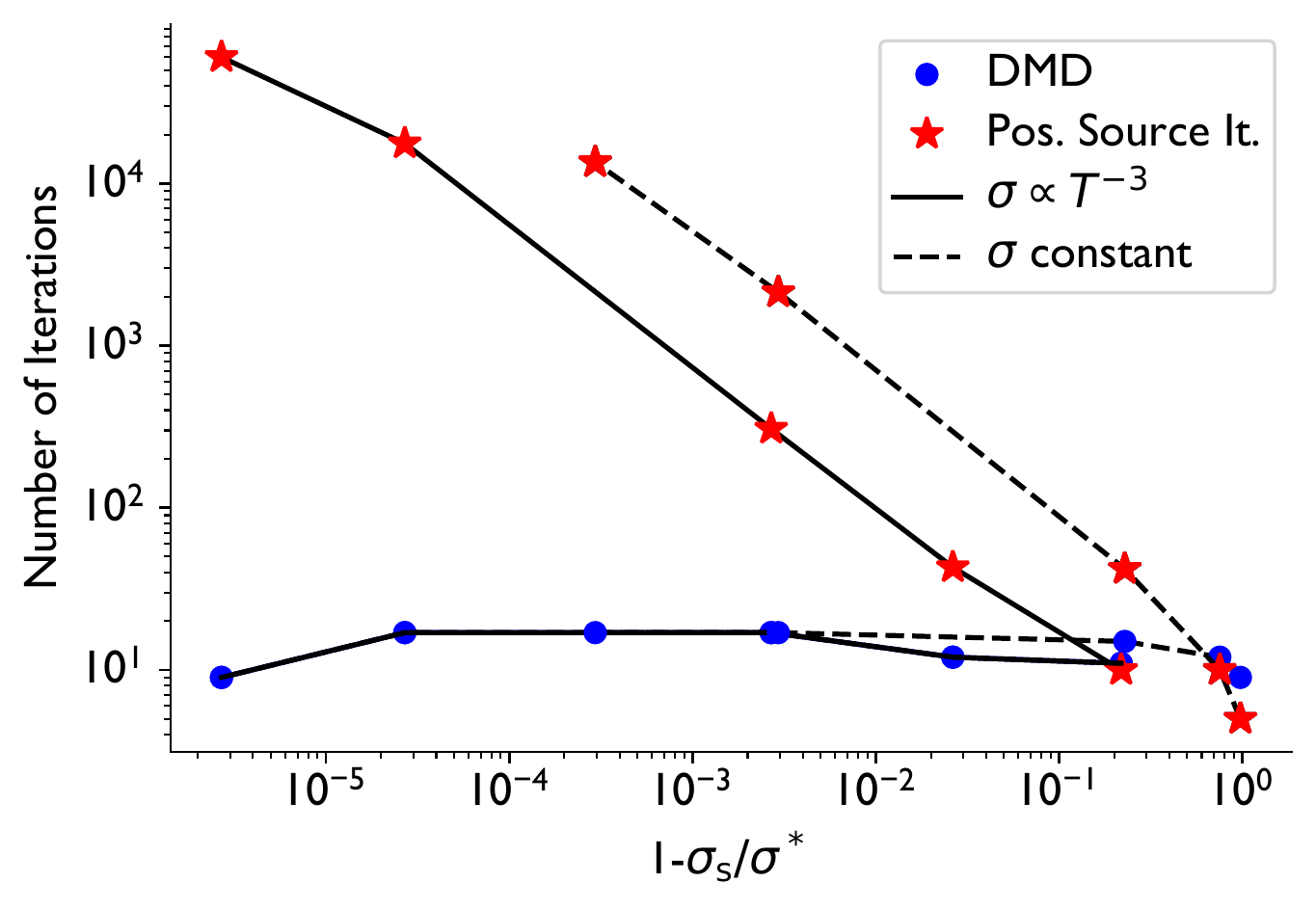}
    \caption{\change{Number of iterations required to solve the cooling problem as a function of the effective scattering ratio.}}
    \label{fig:cooling_converge}
\end{figure}

\change{
The number of iterations required to solve this problem is shown in Figure \ref{fig:cooling_converge}. In the figure we include results for both forms of the opacity to demonstrate that added nonlinearity does not seem to affect the behavior. The problem has the scattering ratio range from about 0.8 up to $1-2.7 \times 10^{-6}$. The number of iterations required for the DMD-accelerated method is roughly constant  as the scattering ratio approaches unity. This is clearly not the case for positive source iteration as the number iterations grows rapidly as the scattering ratio approaches one.  Indeed, we had to terminate the highest scattering ratio case at $6\times 10^4$ iterations before reaching convergence. In this hardest case, there number of iterations is smaller by a factor of over 6000 for DMD.}

\begin{figure}
    \centering
    \subfigure[Piecewise Cubice Solution]{\includegraphics[width=0.49\textwidth]{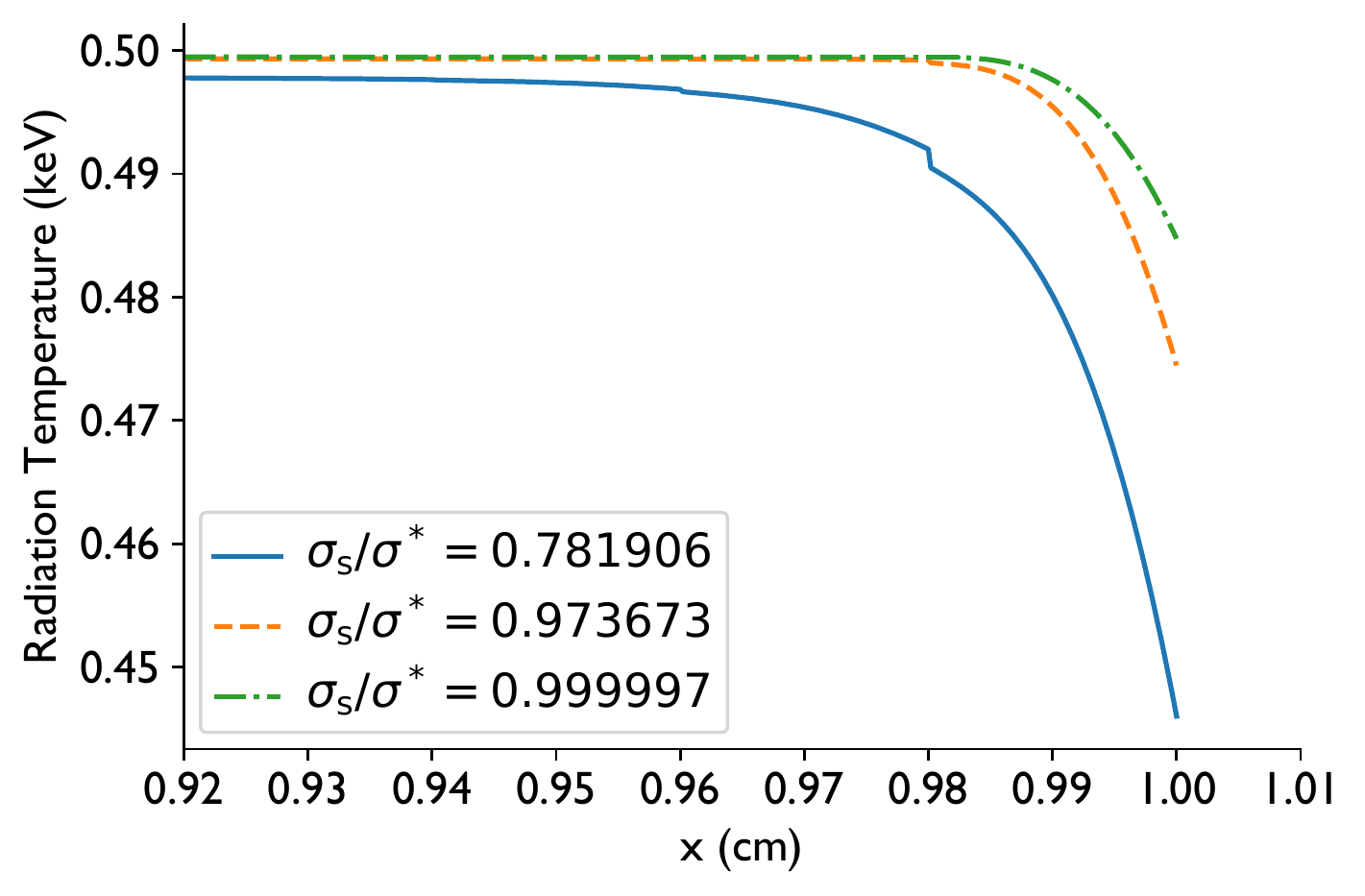}}
    \subfigure[Singular values of $Y^-$]{\includegraphics[width=0.49\textwidth]{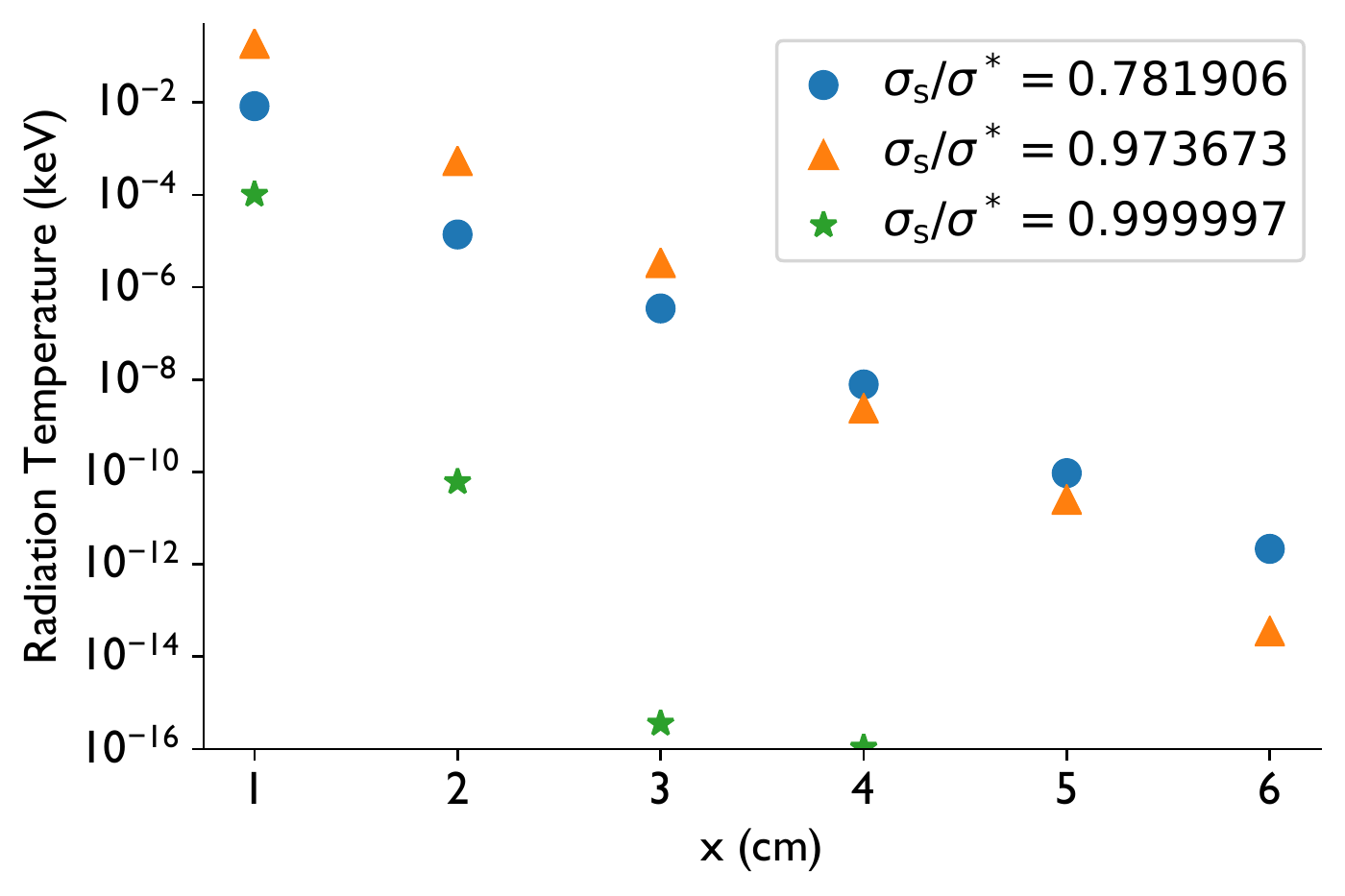}}
    \caption{\change{The solution near the slab edge in the cooling problem  and the singular values of the $Y^-$ data matrix  as a function of the effective scattering ratio.}}
    \label{fig:cooling_structure}
\end{figure}

\change{This problem demonstrates that the effectiveness of DMD as the limit as the scattering ratio approaches unity. The results also demonstrate that as the scattering ratio increases, it is possible for DMD to have the number of iterations go down.  We have observed this in other problems of linear particle transport. We believe that this is due to there being less spatial structure in the solution as the scattering ratio approaches one because there is less cooling of the block. As shown in Figure \ref{fig:cooling_structure}(a) the solution has less spatial detail as the scattering ratio approaches one. With a lower scattering ratio there is a discontinuity at the cell edge near 0.98 cm that diminishes as the scattering ratio increases. Further evidence for this effect comes from the singular values of the data matrix $Y^-$ displayed in Figure \ref{fig:cooling_structure}(b).  For the time step leading to a scattering ratio nearest to unity, there are only two singular values larger than $10^{-14}$.  This means that the data can be well represented by a rank 2 approximation. This is not the case in the two lower scattering ratio cases in Figure \ref{fig:cooling_structure}. Both of these cases have at least 6 singular values above $10^{-14}$ and, therefore, have more structure in the solution that DMD needs to approximate. This leads to more iterations being needed to form the DMD approximation. Our algorithm using the sequential SVD detects that this is the case and requires fewer iterations when the rank of the data matrix is lower.
}

\subsection{Su-Olson Test}
There also exist semi-analytic solutions for radiative transfer in an optically thin problem, driven by a radiation source where the heat capacity of the material is proportional to the cube of the temperature. Transport and diffusion solutions for this problem can be found in \cite{Su19971035} and S$_2$ solutions\footnote{The solutions are given for the P$_1$ equations, but S$_2$ with Gauss quadrature is equivalent to P$_1$ in slab geometry.}  are given in \cite{mcclarren2008analytic}. We solve this problem to demonstrate that the DMD-accelerated solution is not slower to converge than positive source iteration in optically thin media, where we would expect that no acceleration is needed. 

\begin{figure}
    \subfigure[Radiation]{\includegraphics[width=0.49\textwidth]{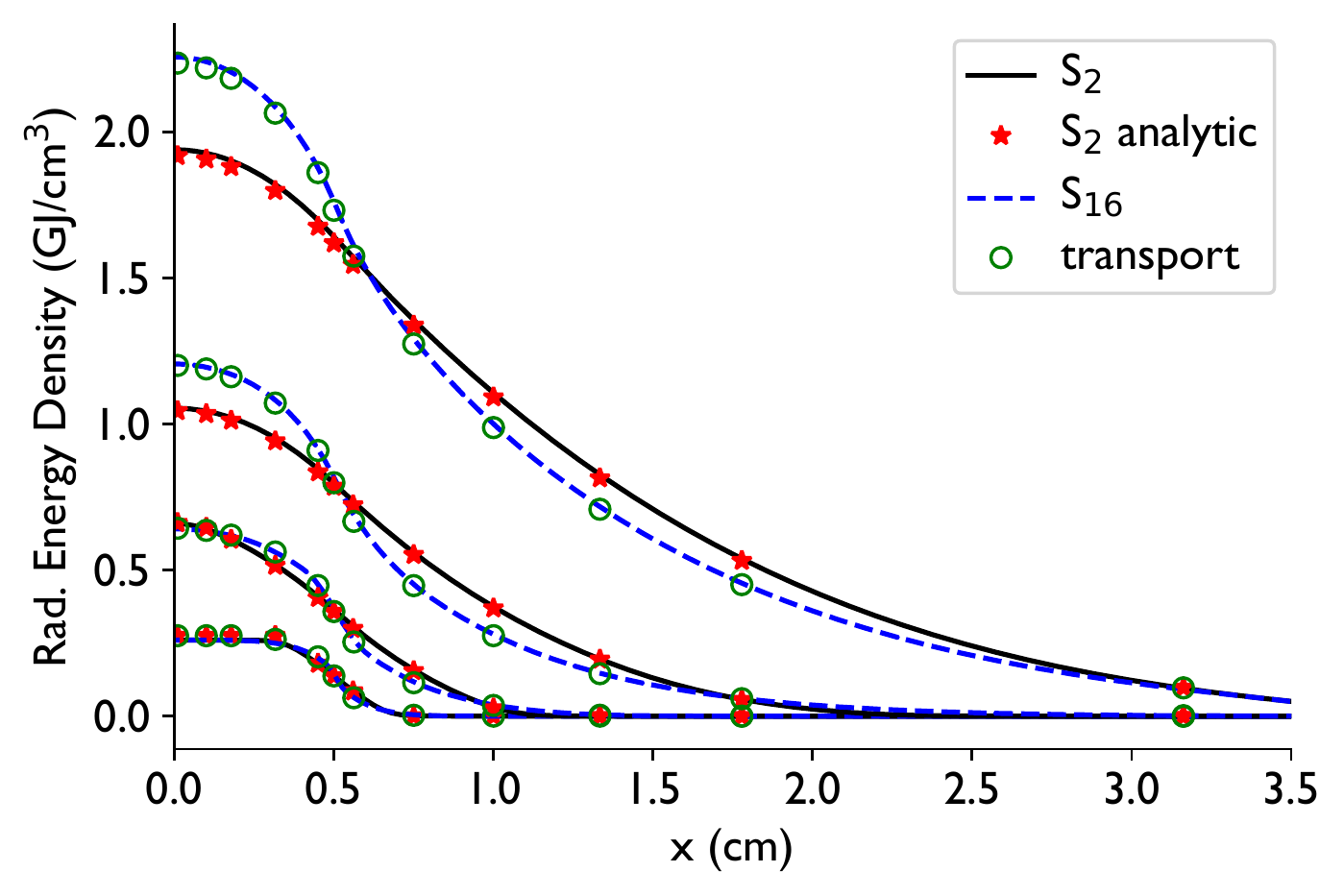}}
    \subfigure[Material Temperature]{\includegraphics[width=0.49\textwidth]{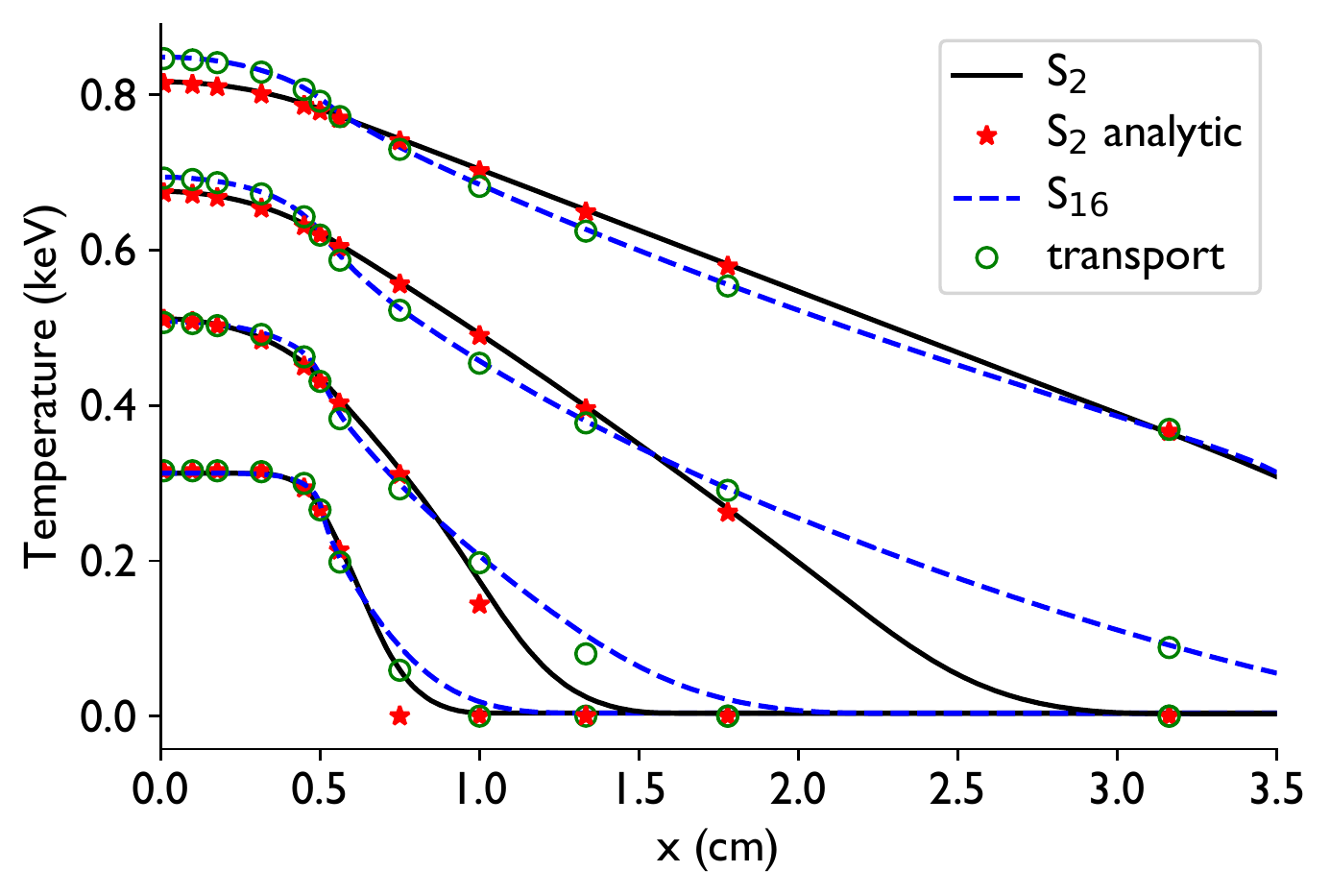}}
    \caption{Comparison of the numerical results from DMD-accelerated S$_N$ and the S$_2$/P$_1$ and transport analytic solutions at times $t= 0.316228 \sigma/c, \sigma/c, 3.16228 \sigma/c,$ and $10\sigma/c$ with $\sigma=1\,\mathrm{cm}^{-1}$.}
    \label{fig:suRadProb}
\end{figure}

For this problem we observe that the number of iterations required is almost identical between DMD and source iteration. At most we see a 5\% decrease in the number of iterations per time step. 
A comparison of the numerical and analytic solutions are shown in Figure \ref{fig:suRadProb}.

\subsection{Laser-Driven Radiating Shock Problem}
The final problem we solve is a radiative transfer problem inspired by experiments involving laser-driven shocks \cite{holloway2011predictive}. In these experiments a laser pulse strikes a beryllium (Be) disk that is on the end of a xenon (Xe) filled tube. The laser launches a shock wave into the Be disk that eventually breaks out into the Xe gas.  The state of the system at a given time \cite{mcclarren2011physics} is used to set up our test problem.   The radiative transfer in these shock experiments is complicated due to the large thin sources that arise \cite{MacDrake,McClarrenShock,drake2007,holgado2015anti}.  This problem will test how the DMD acceleration technique performs on a realistic problem with multiple materials, large variations in density, and optically thin and thick regions. \change{The optically thick regions of this problem necessitate acceleration of positive source iteration (as we will show below).}

The density, temperature, and material (either Be or Xe) for the test problem are given in Table \ref{tab:beshock}. This table gives the initial conditions for the temperature and the intensity in equilibrium (i.e., $I = acT^4/2$). The boundary conditions assume an incoming, isotropic source corresponding to the temperature nearest the boundary.
\begin{table}[]
    \centering
 \begin{tabular}{cccc}
 Position (mm) & Density (g/cm$^3$) & Temperature (eV) & Material\\
\hline
 0.0000 & 0.0168 & 40.1723 & Be \\
 0.0462 & 0.1681 & 11.6676 & Be \\
 0.1046 & 0.3429 &  5.5016 & Be \\
 0.1080 & 0.5107 &  3.6279 & Be \\
 0.1134 & 0.7383 &  1.9934 & Be \\
 0.1241 & 0.1829 &  4.6652 & Be \\
 0.1300 & 0.1384 &  6.2204 & Be \\
 0.1302 & 0.7405 & 16.2775 & Xe \\
 0.1322 & 0.0493 & 74.5469 & Xe \\
 0.6000 & 0.0065 & 14.9126 & Xe \\
\hline
\end{tabular}
    \caption{Density, initial temperature, and material as a function of position for the radiating shock problem.}
    \label{tab:beshock}
\end{table}
Between the points in the table we linearly interpolate to evaluate the density and initial temperature.  All points to the left of $0.1302$ cm are beryllium, and the remainder is xenon.
\begin{figure}
    \centering
    \includegraphics[width=0.49\textwidth]{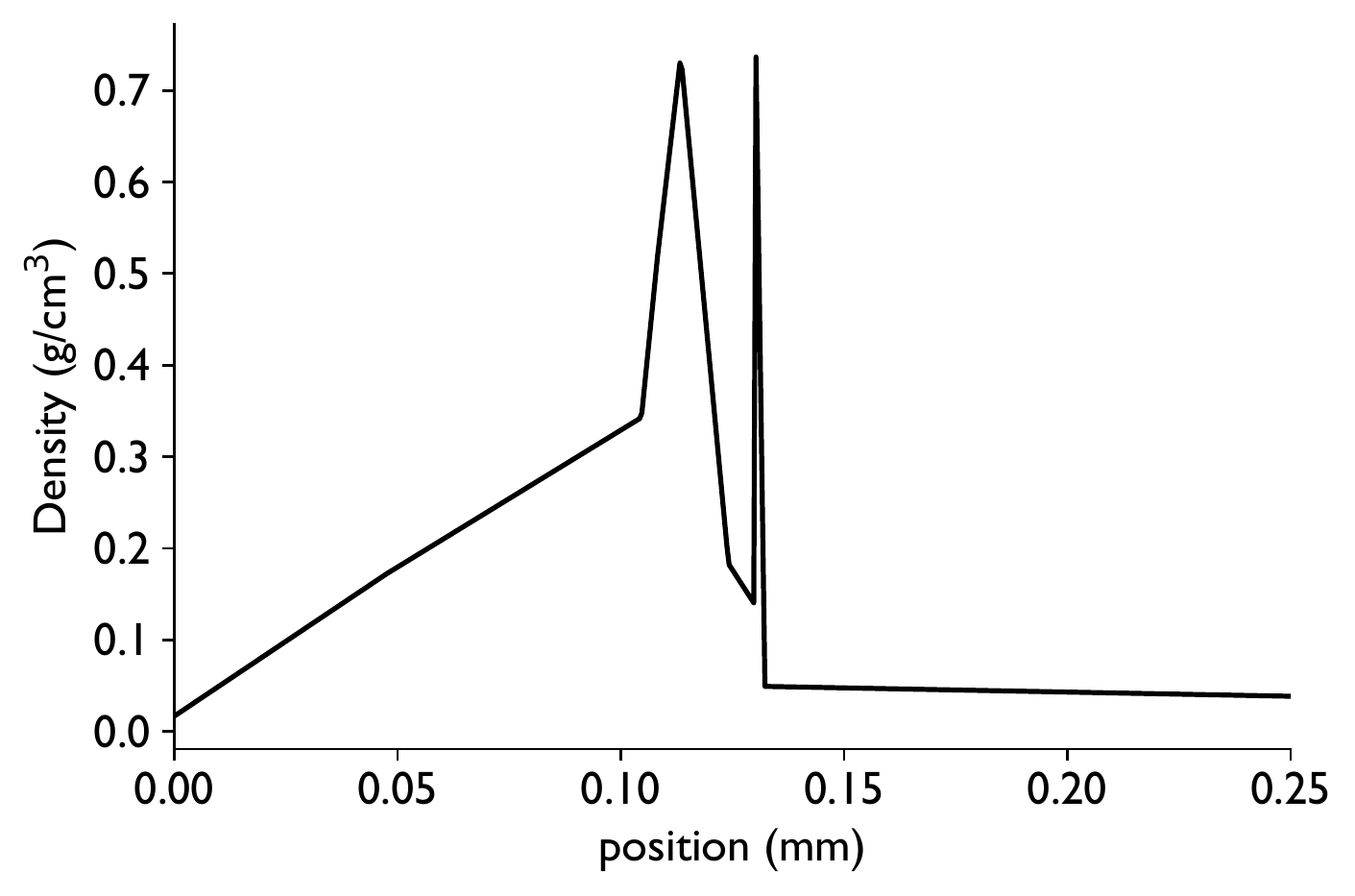}
    \includegraphics[width=0.49\textwidth]{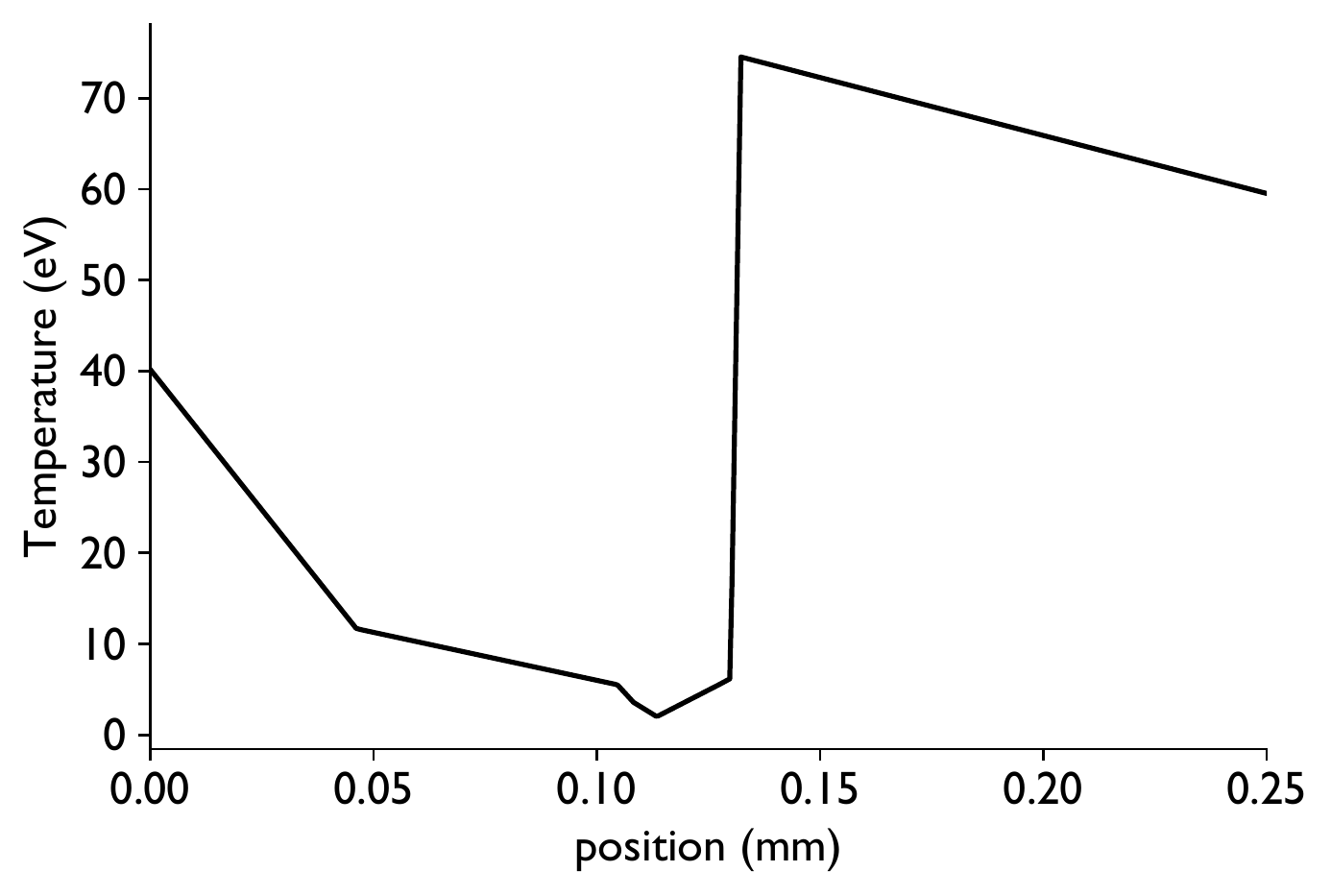}
    \includegraphics[width=0.49\textwidth]{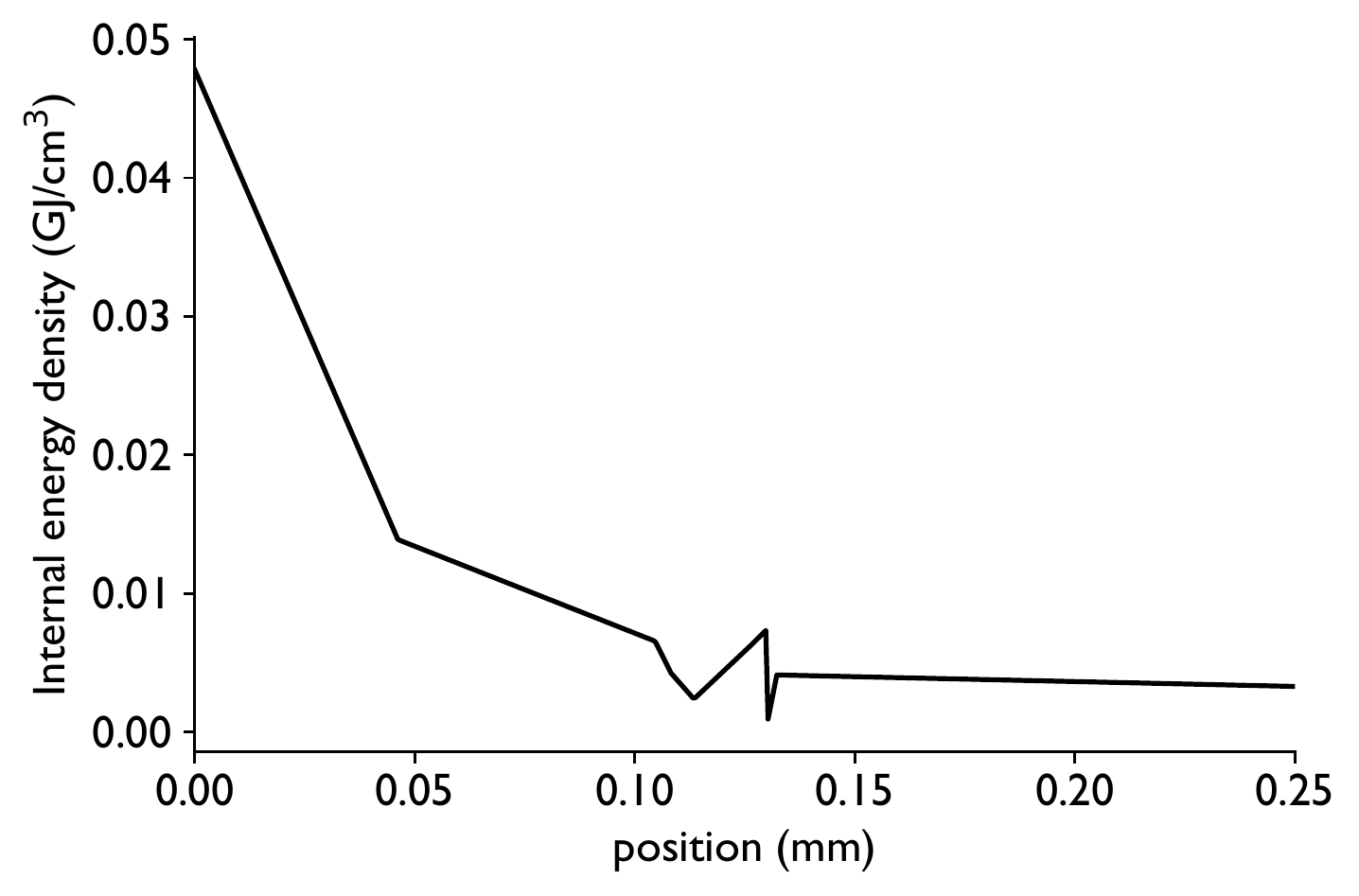}
    \includegraphics[width=0.49\textwidth]{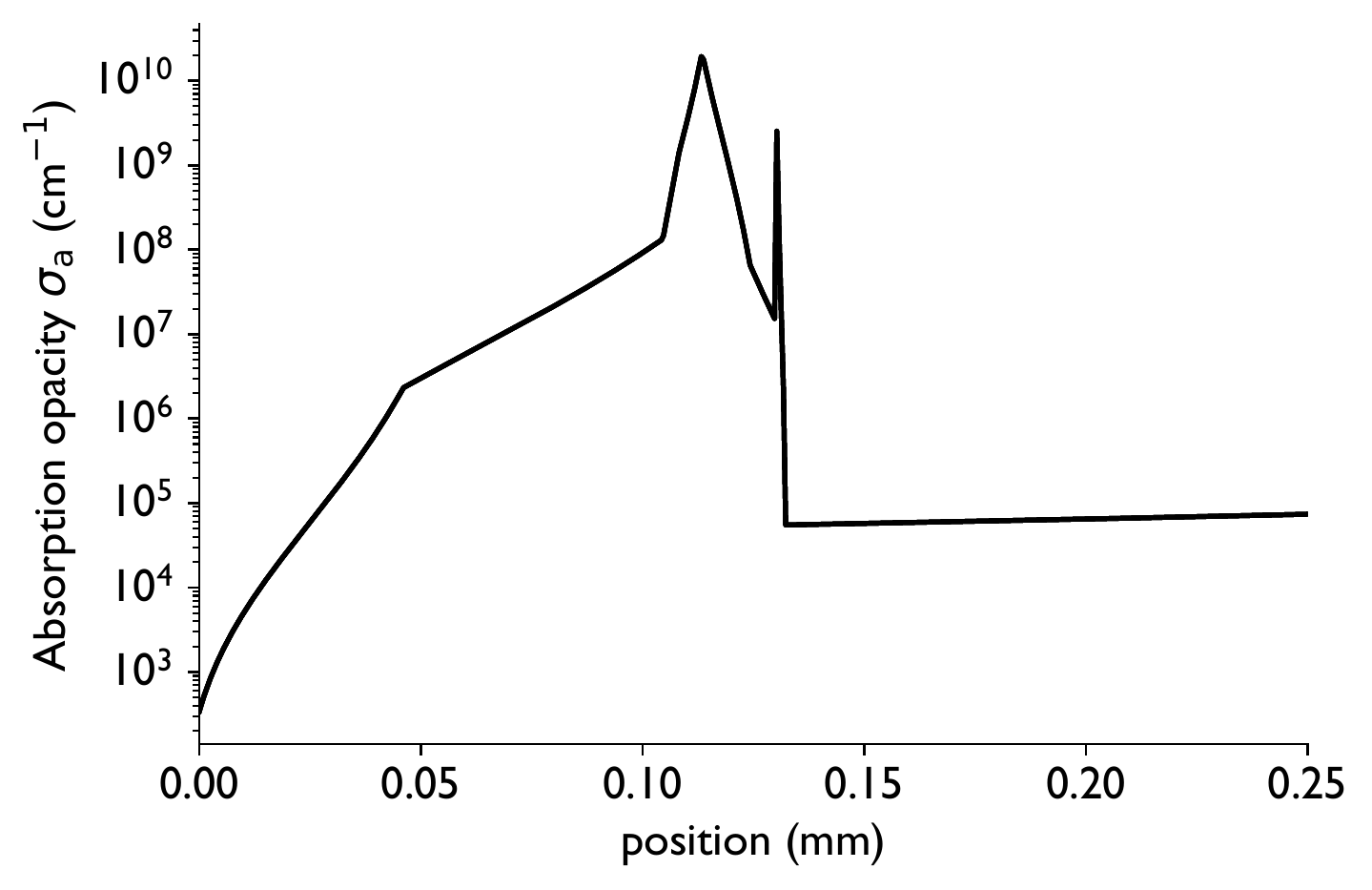}
    \caption{The density and initial temperature, internal energy density, and $\sigmaa$ for the radiating shock problem. }
    \label{fig:radshock_int}
\end{figure}

The heat capacities are based on a gamma-law equation of state with $\gamma = 5/3$ in xenon and $\gamma = 1.45$ in beryllium as calibrated from experiment \cite{stripling2013calibration,chakraborty2013spline} to give
\begin{equation}
    C_\mathrm{v} \left[ \frac{\mathrm{GJ}}{\mathrm{keV}\cdot\mathrm{cm}^{3}} \right] = \begin{cases}
    1.1899 & \text{in Be} \\ 0.05513 & \text{in Xe}
    \end{cases} .
\end{equation}
Additionally, we use an approximate bremsstrahlung opacity \cite{ZR} as
\begin{equation}
    \sigmaa(T)\left[ \mathrm{cm}^{-1}\right]  = 
    0.088 \rho^2 Z^2 T^{-\frac{7}{2}},
\end{equation}
for $T$ in keV, $\rho$ the density in g/cm$^3$, and $Z$ is the atomic number of the material, $4$ for Be and $54$ for Xe.

In Figure \ref{fig:radshock_int} the density and the initial values for the temperature, internal energy density, and $\sigmaa$ are shown as a function of position. At the Be/Xe interface there is a jump in the temperature due to the fact that the two materials have different heat capacities. Between this interface and the other density maximum at 0.1134 cm there is a region where the absorption opacity drops. This is where most of temperature change due to radiative transfer in this problem will occur. Due to the stiffness of the problem from the large opacity and small value for the heat capacity, we use the modified linearization from \cite{McClarren:2009p1640} with $\ell = 5$. This has the effect of reducing the value of $f$ in Eq.~\eqref{eq:fleck} and increasing the scattering.

\begin{figure}
    \centering
\includegraphics[width=0.8\textwidth]{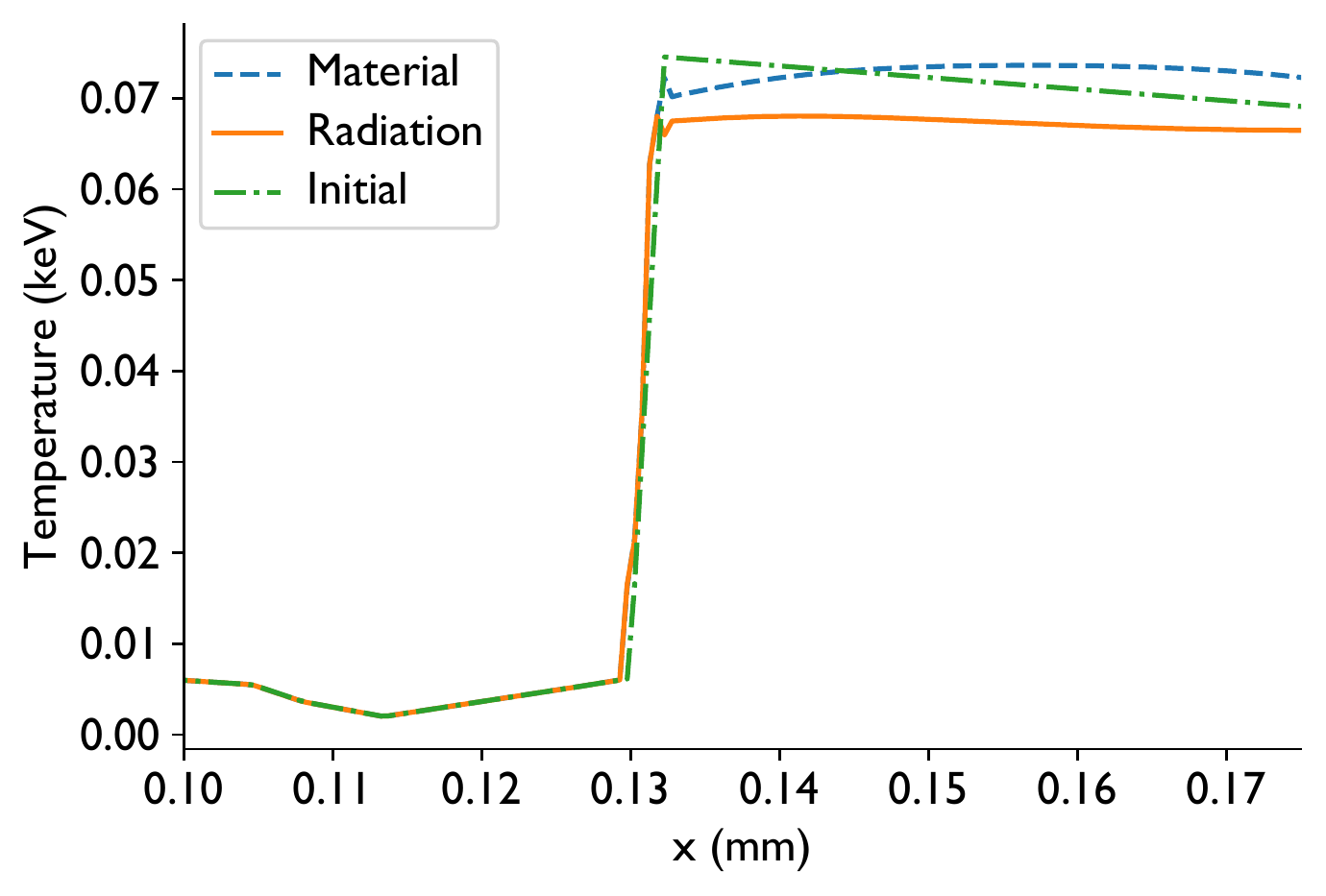}
    \caption{Solution from DMD-accelerated S$_{12}$ for the radiating shock problem at $t=1$ ns along with the initial condition.}
    \label{fig:radshocksol}
\end{figure}

To compare the efficiency of our DMD acceleration we solve the radiative transfer problem for this shock profile over a time step of $0.01$ ns. We consider a spatial domain extended from $x=0$ to $0.25$ mm with 500 spatial zones and use order 3 finite elements. The DMD-accelerated solution required 42 iterations while positive source iteration required 827; the acceleration led to a speed up of nearly a factor of 20 times. In other words the accelerated solution could complete 20 time steps for the cost of a single, unaccelerated time step.  The solution for this problem after 100 time steps (i.e., $t=1$ ns), is shown in Figure \ref{fig:radshocksol}.

\section{Conclusions and Future Work}
We have presented a novel method for accelerating the discrete ordinates solution of radiative transfer problems that includes nonlinear positivity preservation. The acceleration technique, based on the dynamic mode decomposition and an incremental singular value decomposition, was found to be a factor of nearly 20 faster than positive source iteration on a radiative shock problem and above three times faster for a standard Marshak wave problem.

There is clearly more research that should be performed on this method.  For instance, we used a linearization of the radiative transfer equations and did not consider nonlinear temperature updates or non-gray radiative transfer.  Using DMD as a part of multigroup, nonlinear elimination, as outlined in \cite{brunnerElimination}, could be a fruitful avenue of investigation.  There are other radiative transfer problems that could also benefit from a DMD approach.  For instance, the iterative implicit Monte Carlo (IIMC) method of Gentile and Yee \cite{gentile2016iterative} requires a series of iterations similar to source iterations.  It is possible that DMD acceleration could also perform well on that formulation.  

\change{Outside of radiative transfer there is other important future work that could be explored.  DMD could be used in linear particle transport problems (e.g., neutron transport) to accelerate diffusion synthetic accelerated schemes (DSA) or GMRES iterations by wrapping the iterations in a DMD-like approach.  Additionally, for linear problems a comparison, in terms of memory, runtime, etc., between DMD and GMRES would be appropriate because both approaches store information about the iterations to approximate a solution. }


\section*{ACKNOWLEDGMENTS}
\change{The authors would like to acknowledge the anonymous referees for their suggestions to improve this paper and thought-provoking questions.}

Lawrence Livermore National Laboratory is operated by Lawrence Livermore National Security, LLC, for the U.S. Department of Energy, National Nuclear Security Administration under Contract DE-AC52-07NA27344.  This document (LLNL-JRNL-813575) was prepared as an account of work sponsored by an agency of the U.S. government.  Neither the U.S. government nor Lawrence Livermore National Security, LLC, nor any of their employees makes any warranty, expressed or implied, or assumes any legal liability or responsibility for the accuracy, completeness, or usefulness of any information, apparatus, product, or process disclosed, or represents that its use would not infringe privately owned rights.  Reference herein to any specific commercial product, process, or service by trade name, trademark, manufacturer, or otherwise does not necessarily constitute or imply its endorsement, recommendation, or favoring by the U.S. government or Lawrence Livermore National Security, LLC. The views and opinions of authors expressed herein do not necessarily state or reflect those of the U.S. government or Lawrence Livermore National Security, LLC, and shall not be used for advertising or product endorsement purposes. 

\bibliography{radtran}

\end{document}